\documentclass[12pt,preprint]{aastex}
\def\gtorder{\mathrel{\raise.3ex\hbox{$>$}\mkern-14mu
                \lower0.6ex\hbox{$\sim$}}}
\def\ltorder{\mathrel{\raise.3ex\hbox{$<$}\mkern-14mu
                \lower0.6ex\hbox{$\sim$}}}

\shorttitle{Photoionized Gas in Dark-Matter Minihalos}
\shortauthors{Gnat \& Sternberg}

\begin{document}
\title{Photoionized Gas in Dark-Matter Minihalos in
the Galactic Halo and Local Group}
\vspace{1cm}
\author{Orly Gnat and Amiel Sternberg}
\vspace{0.5cm}
\affil{School of Physics and Astronomy and the Wise Observatory,
        The Beverly and Raymond Sackler Faculty of Exact Sciences,
        Tel Aviv University, Tel Aviv 69978, Israel}
\email{orlyg@wise.tau.ac.il}

  
\begin{abstract}
We present computations of the metals photoionization structures
of pressure supported gas clouds in gravitationally dominant
dark-matter mini-halos
in the extended Galactic halo or Local Group environment. We consider
low-metallicity ($0.1-0.3$ solar)
clouds that are photoionized by the present-day 
metagalactic radiation field, and which are also 
pressure-confined by a hot external medium. We study the
properties of $\Lambda$CDM Burkert (and also NFW) 
mini-halos with characteristic circular velocities from 12
to 30~km~s$^{-1}$ ($10^8$ -- $2\times 10^9$ $M_\odot$).
We present results for the volume densities and projected column
densities of low-ionization metal atoms and ions, 
\ion{C}{2}, \ion{N}{1}, \ion{O}{1}, \ion{Si}{2}, and \ion{S}{2},
produced in the neutral cloud cores, and high-ions, 
\ion{C}{4}, \ion{N}{5}, \ion{O}{6}, \ion{Si}{3}, \ion{Si}{4}, and \ion{S}{3}
produced in the ionized shielding envelopes.
We examine the possible relationships between the 
compact \ion{H}{1} high-velocity clouds (CHVCs), dwarf galaxies, and
high-velocity ionized \ion{C}{4} absorbers, within the context
of photoionized mini-halo models for such objects.
In pressure-confined ($P/k \sim 50$ cm$^{-3}$ K)
mini-halo models for the CHVCs
the photoionization states are much lower than in the
\ion{C}{4} absorbers. However, for lower bounding pressures 
$\lesssim 1$ cm$^{-3}$ K, in more massive
$\sim 2\times 10^9$ $M_\odot$ dwarf galaxy scale halos,
the photoionization states in the outer
envelopes resemble those observed in the \ion{C}{4} HVCs.
An important exception is \ion{O}{6} for which our
photoionization models underpredict the relative abundances
by about an order-of-magnitude,
suggesting that additional processes are at play as concluded
by Sembach et al.~(2000; 2003).
For the cloud size-scales expected in median
$\Lambda$CDM Burkert halos, a gas metallicity
of $0.3$ solar is required to produce absorption-line
strengths
comparable to those in the \ion{C}{4} HVCs.
We argue that fully ionized and starless ``dark galaxies''
could be detectable in the local universe as
UV metal line absorbers with ionization states
similar to the \ion{C}{4} absorbers.

\end{abstract}

\keywords{dark matter -- Galaxy:evolution -- Galaxy:formation
-- Galaxy:general -- intergalactic medium -- Local Group
-- ISM:abundances -- quasars:absorption lines}


\section{Introduction}
\label{introduction}

Recent ultraviolet (UV) absorption-line observations have revealed the
existence of highly ionized ``high-velocity'' gas
clouds along lines of sight to extragalactic UV continuum sources
(Sembach et al.~1999, 2000, 2002, 2003; Murphy et al.~2000;
Wakker et al.~2003).
These gas clouds may be the ionized components, or 
counterparts, of the well studied 
neutral hydrogen (\ion{H}{1}) high-velocity
clouds (HVCs) (Muller, Oort \& Raimond 1963;
Wakker \& van Woerden 1997; Putman et al.~2002; 
de Heij, Braun \& Burton 2002), objects which are
usually defined as \ion{H}{1} systems with radial velocities inconsistent
with differential rotation in the Galactic disk.
The origin and nature of the HVCs remains controversial. 
One attractive possibility is that at least some of
the high velocity gas, both neutral and ionized,
represents remnant ``infalling'' material
associated with galaxy formation processes
(Oort 1970; Blitz et al.~1999; Braun \& Burton 1999;
Sternberg, McKee \& Wolfire 2002). 

In the UV studies, many high-ionization 
metal absorption lines have been detected,
including \ion{Si}{3} $\lambda~1206.5$, \ion{Si}{4}
$\lambda\lambda~1393.8, 1402.8$, 
and \ion{C}{4} $\lambda\lambda~1548.2, 1550.8$, 
observed with GHRS and STIS
on board HST, 
and more recently \ion{O}{6} $\lambda 1031.9$ absorption
observed with FUSE. Some of the ionized UV absorbers can, in fact,
be identified with the large ($\sim 100$ deg$^2$)
and nearby ($\lesssim 50$ kpc)
\ion{H}{1} HVC ``complexes'', such as
the Magellanic Stream, or Complex C (Sembach et al.~2003).
However, some absorbers, such as those observed towards
Mrk 509 and PKS 2155-304 (Sembach et al.~1999; Collins, Shull \& Giroux~2004) 
do not appear directly linked
to such structures, and could be clouds at much larger
distances, perhaps distributed within the Local Group.
The ``high-ion absorbers'' may be
the local analogs of metal-line absorbers observed
in damped Ly$\alpha$ systems at high redshifts 
(Lu et al.~1996; Prochaska \& Wolfe~1997;
Boksenberg, Sargent \& Rauch~2003;
Maller et al. 2003).

Sembach et al.~(1999) suggested that photoionization by metagalactic 
radiation in low density clouds may be responsible for the \ion{C}{4}
observed towards Mrk~509 and PKS~2155-304. However, the detection of highly
ionized \ion{O}{6} along these sight-lines strongly suggests that
additional processes, such as collisional ionization in conductive 
interfaces or in thermally unstable cooling gas,
are important (Heckman et al.~2002; Fox et al.~2003). 

In this paper we reexamine the properties of purely photoionized
clouds, within the context of dark-matter ``mini-halo'' models
for the compact \ion{H}{1} high-velocity clouds (CHVCs) and dwarf galaxies.
The CHVCs are of particular interest. They are 
kinematically distinct objects, with
typical angular diameters  $\lesssim 1^\circ$
as determined by the recent HIPASS and Leiden-Dwingeloo 21 cm surveys
(Putman et al.~2002; de Heij et al.~2002),
and they may be at larger characteristic distances from the Galaxy
disc than are the large HVC complexes.
In Sternberg, McKee \& Wolfire (2002; hereafter SMW) we
explored the hypothesis (Blitz et al.~1999; Braun \& Burton 1999)
that the population of CHVCs traces dark-matter (DM) substructure
(Moore et al.~1999; Klypin et al.~1999)
in the Galactic halo or in the more extended Local Group 
environment.  
In this picture the CHVCs are essentially
the lower-mass ``cousins'' of the Local Group dwarf galaxies,
and represent objects in which gas is confined to 
low-mass dark-matter ``mini-halos''. In SMW we focused on
the phase properties of hydrogen gas, ionized/neutral/warm/cold, of
hydrostatic gas clouds in gravitationally dominant
DM halos, including the effects of photoionization
and heating by the metagalactic radiation field, and pressure
confinement by an external hot medium. 
Explicit models were constructed for the observed distributions of warm 
\ion{H}{1} in the Local Group galaxies Leo~A and Sag~DIG. The \ion{H}{1}
boundaries in these objects are consistent with a cutoff due to 
photoionization by the metagalactic field.

The aim of this paper is
to study the associated metals photoionization structures
and resulting absorption-line properties
of such ($10^8 - 2\times 10^9$ $M_\odot$) mini-halo clouds. 
We consider photoionization by the present-day metagalactic field, which 
is the dominant source of radiation at distances $\gtorder~150$~kpc
from the Galactic disk (SMW).
The calculations and models we
construct are similar to those presented by Kepner et al.~(1999)
in their mini-halo models
for the QSO absorption-line systems. Our models are more directly
applicable to CHVCs in the outer Galactic halo
and dwarf galaxies
in the Local Group. In particular, we determine the
conditions required for the formation of highly ionized
absorbers similar to the \ion{C}{4} HVCs.  
 

In \S2 we discuss the basic parameters of our mini-halo models, and describe
the methods we used to compute the metals photoionization structure.
In \S3 we present auxiliary
computations of the abundances of highly ionized metal species
in uniform density and optically thin clouds 
that are photoionized by the present-day metagalactic radiation field.
In \S4 we present photoionization calculations for
mini-halo clouds for a range of
dark-matter and gas masses, metallicities, 
radiation fields and bounding gas pressures.
We compare our results to the \ion{C}{4} 
absorbers towards Mrk~509 and PKS~2155-304. We summarize in \S5.


\section{Model Ingredients}

We are interested in studying the 
metals photoionization structures
of mini-halo clouds similar to those considered by SMW in their
analysis of the \ion{H}{1} CHVCs and Local Group dwarf galaxies.
Following SMW, we construct models for warm, $T=10^4$ K,
pressure supported, spherically symmetric
clouds that are; (a) in hydrostatic
equilibrium in gravitationally dominant dark-matter halos,
(b) photoionized by the present-day
metagalactic radiation field, and (c) subjected to an
external bounding pressure provided by an external hot ionized
medium (HIM). The clouds typically consist of
``warm neutral medium'' (WNM) cores,
surrounded by ``warm ionized medium'' (WIM) 
shielding envelopes that absorb the incident photoionizing radiation.
The WNM/WIM mass ratios depend on 
several quantities, including the total (ionized plus neutral)
warm cloud masses $M_{\rm WM}$, the depths of the dark-matter gravitational
potential wells, the bounding HIM pressures, and the intensity
of the ionizing radiation. In SMW we determined the 
conditions required for the formation of multi-phased mixtures
of ``cold neutral medium'' (CNM) condensations
together with the WNM in the
cloud cores. In this paper we present
calculations of the metals photoionization structures.
We use our models to compute the distributions and
column densities of metal ions in mini-halo clouds for a range
of conditions. 

For the dark-matter densities, we 
focus mainly on halos with the empirically based
Burkert (1995) profiles
\begin{equation}
\rho_{\rm DM} = \frac{\rho_{ds}}{(1+x)(1+x^2)}
\end{equation}
where $x\equiv r/r_{s}$ is the radial distance from the
halo center measured relative to the
scale radius $r_s$, and $\rho_{ds}$ is the characteristic
dark-matter scale density. We also consider one NFW model
(Navarro, Frenk, \& White~1995), for which
\begin{equation}
\rho_{\rm DM} = \frac{\rho_{ds}}{x(1+x)^2} \ \ \ .
\end{equation}
Cosmological simulations and observations
suggest that in realistic halos
the scale parameters $r_s$ and $\rho_{ds}$
\footnote{In Burkert halos, $\rho_{ds}$ is the dark-matter
density at $x=0$. For NFW halos $\rho_{ds}$
is the density at $x=0.46$.} are correlated.  
We adopt a $\Lambda$CDM ($H_0 = 0.70$~km~s$^{-1}$~Mpc$^{-1}$, 
$\Omega_m = 0.3$, $\Omega_{\Lambda} = 0.7$ and
$\sigma_8 = 0.9$) correlation as specified by the formula
\begin{equation}
x_{\rm vir} = 27\times 10^{0.14\sigma}
\biggl(\frac{M_{\rm vir}}{10^9 \ M_\odot}\biggr)^{-0.08}
\end{equation}
that relates the concentration parameter $x_{\rm vir}\equiv r_{\rm vir}/r_s$
(where $r_{\rm vir}$ is the virial radius) and virial mass
$M_{\rm vir}$, for low-mass halos 
(Bullock et al. 2001; SMW; Wechsler et al.~2002).
In this expression
$\sigma$ is the standard deviation
for underconcentrated or overconcentrated halos.
Given $M_{\rm vir}$ and $x_{\rm vir}$, the scale
density and radius are given by 
$\rho_{ds}~=~\Delta\rho_ux_{\rm vir}^3/f_M(x_{\rm vir})$
and
$r_s~=~(3/4\pi\Delta\rho_u)^{1/3}M_{\rm vir}^{1/3} /x_{\rm vir}$,
where $\rho_u$ is the mean cosmic matter density, $\Delta\sim 340$
is the mean overdensity in virialized halos, and $f_M(x_{\rm vir})$
is the dimensionless DM mass enclosed within the virial radius (SMW).
The characteristic halo scale velocity is defined by 
$v_s^2\equiv (4\pi G/3)\rho_{ds}r_s^2$. 

For the (isotropic) metagalactic radiation field we 
consider two options. First, we assume a ``standard field'' as
given by the SMW fit
\begin{equation}
\label{radn}
J_{\nu} = \left\{ \begin{array}{lcl}
   1.051\times10^2J_{\nu0}(\frac{\nu}{\nu_0})^{-1.5}&,& \frac{\nu}{\nu_0}~<~0.3\\
   J_{\nu0}(\frac{\nu}{\nu_0})^{-5.41}&,& 0.3~<~\frac{\nu}{\nu_0}~<~1\\
   J_{\nu0}(\frac{\nu}{\nu_0})^{-3.13}&,& 1~<\frac{\nu}{\nu_0}~<~4\\
   2.512\times10^{-2}J_{\nu0}(\frac{\nu}{\nu_0})^{-0.46}&,& 4~<~\frac{\nu}{\nu_0} \\
                  \end{array}\right.
\end{equation}
for the 1.6 $\mu$m to 3 keV range.  In this expression
$J_{\nu0} = 2\times10^{-23}$ erg s$^{-1}$ cm$^{-2}$ Hz$^{-1}$ sr$^{-1}$,
and $\nu_0$ is the Lyman limit frequency. This fit, which we
plot in Figure 1, is based on
observational constraints for the optical/UV and X-rays
(Bernstein, Freedman \& Madore 2002; Martin, Hurwitz \& Bowyer 1991;
Chen, Fabian, \& Gendreau 1997).
 For the unobservable radiation from the Lyman limit to $\sim0.25$~keV
it is based on theoretical models 
(Haardt \& Madau 1996; see also Shull et al.~1999).
Second, we consider a ``maximal field'' that is enhanced
between the Lyman limit and 0.25 keV as illustrated by
the dashed line in Figure 1.  For photon energies above
the Lyman limit our maximal field is represented by
\begin{equation}
\label{radm}
J_\nu = \left\{ \begin{array}{lcl}
   J_{\nu0}(\frac{\nu}{\nu_0})^{-1.725}&,& 1~<\frac{\nu}{\nu_0}~<~18.4\\
   2.512\times10^{-2}J_{\nu0}(\frac{\nu}{\nu_0})^{-0.46}&,& 18.4~<~\frac{\nu}{\nu_0} \\
                  \end{array}\right.
\end{equation}
It is identical to the standard field in the optical/UV and
X-rays, except for the enhancement 
from the Lyman limit to $\sim0.25$~keV.
The Lyman continuum photon intensities
implied by expressions (\ref{radn}) and (\ref{radm}) are both
consistent with measured upper-limits on the H$\alpha$ surface brightnesses
of intergalactic clouds (Vogel et al.~1995; Weiner et al.~2001).

In our numerical computations, we begin by calculating the coupled 
hydrostatic and ionization structures of pure hydrogen-helium mini-halo clouds,
assuming a primordial helium abundance $n_{\rm He}/n_{\rm H} = 1/12$, and
following the iterative radiative transfer
procedures outlined by SMW.
For the range of metallicities (0.1 to 0.3 times solar)
and total hydrogen column densities (up to $\sim 10^{21}$ cm$^{-2}$)
that we consider, the opacity
is dominated by the hydrogen and helium absorptions, and metals
are negligible. Dust opacity may also be ignored, assuming
Galactic dust properties, and a dust to gas ratio proportional
to the metallicity
\footnote{At 1000 \AA \
the dust absorption cross section per hydrogen
nucleus $\sigma=1.9\times 10^{-21}$ cm$^{-2}$ for Galactic dust
(Draine~2003). Assuming that the absorption
is proportional to the metallicity, a dust optical depth of $\tau=0.86$
is expected for $N_{\rm H}=1.5\times 10^{21}$ cm$^{-2}$ and
a metallicity of 0.3.}.
The result of these calculations is the
density and ionization profiles of the hydrogen-helium gas, 
extending from the halo center to an outer cloud radius, $r=r_{\rm W/H}$,
where the WIM gas pressure equals the bounding HIM pressure.

Given the \ion{H}{1} density profile we then
calculate the local absorbed radiation field
\begin{equation}
J_\nu(r) = \frac{1}{2} \int _{1}^{-1} J^0_\nu~e^{-\sigma(\nu) N_{{HI}}(r,\mu)}d\mu
\end{equation}
where in this expression $J^0_\nu$ is the unattenuated field, 
$\sigma(\nu) = \sigma_H(\nu)+\sigma_{He}(\nu)/12$ is the
hydrogen-helium photo-absorption cross-section,
$N_{\rm HI}(r,\mu)$ is the
\ion{H}{1} column density from $r$ to the cloud surface 
along a ray inclined at an angle $\theta$ relative to
the radial direction, and where $\mu = cos\theta$.

Finally, given the local hydrostatic hydrogen gas
density, $n_{\rm H}(r)$, and the previously computed
local absorbed radiation field,
$J_\nu(r)$, 
we use the photoionization code
CLOUDY (version 94.00, Ferland et al.~1998
\footnote{See http://www.nublado.org}) 
to calculate the local
equilibrium ionization states for the metals. In this
procedure we typically divide the clouds into up to $800$
optically thin shells. For each thin shell we call CLOUDY
using the local absorbed field computed in our prior
radiative transfer computations.
We assume isothermal 
$10^4$ K gas
\footnote{Our assumed gas temperature
is consistent with explicit CLOUDY computations of the
thermal balance, showing that $T$ lies in the
range of $8\times10^3$ to $3\times10^4$ for the WIM densities
($\lesssim 10^{-2}$~cm$^{-3}$) and
range of metallicities we consider.}
, and fixed metallicity independent of radius.
Given the run of ion volume densities,
we then compute the line-of-sight ion column densities, $N_{\rm ion}(b)$,
where $b$ is the projected distance of the line-of-sight from the
cloud center.    

In \S 4 we present models for a range of conditions. We consider
mini-halos with virial masses from $10^8$ to $2\times 10^9$ $M_\odot$,
as appropriate for the CHVC to dwarf galaxy mass range.
We consider total warm cloud gas masses (ionized plus neutral)
from $10^6$ to $10^9$ $M_\odot$, with
HIM bounding pressures ranging from
0.01 to 50 cm$^{-3}$ K. Our models indicate how
the metal absorption-line properties depend on these
parameters, as well as the ionizing field,
and metallicity.


\section{Photoionization by the Metagalactic Field}

Before we discuss our mini-halo cloud models, we present
auxiliary computations for the ionization states of
uniform density and optically thin gas that is
photoionized by the metagalactic field.
These auxiliary computations will be useful in
understanding the results of our mini-halo cloud models
in which the gas density varies, and in which optical
depth effects are important. Here we
consider individual parcels of $10^4$~K gas with
hydrogen densities $n_{\rm H}$ ranging from
$10^{-7}$ to $1$~cm$^{-3}$. This range corresponds to 
cosmological overdensities
$\delta$ ranging from $\sim1$ to $4 \times 10^6$,
relative to the present-day mean baryon density
$n_b=2.5 \times 10^{-7}$~cm$^{-3}$ 
(for $\Omega_b=0.04$, Spergel et al. 2003).
We assume a metallicity of 0.1 times the solar
abundances as given by Holweger (2001).
We list our 0.1 solar metal
abundances in Table \ref{Holmet}.

In Figure \ref{abun} we plot
\footnote{Color versions of this and other figures appearing
in this paper are available in the electronic on-line edition.},
as functions of $n_{\rm H}$
and for our standard field,
the relative and absolute densities of the various 
``high-ions'',
\ion{C}{4}, \ion{N}{5}, \ion{O}{6}, \ion{Si}{3},
\ion{Si}{4}, and \ion{S}{3},
that we study in \S 4.
We define high-ions as those with ionization thresholds greater
than the hydrogen photoionization threshold of 13.6 eV.
These will generally be present mainly in the
WIM envelopes of our mini-halo clouds.
The upper panel shows the 
{\it fractional} ion abundances, 
$n_{\rm ion}/n_{\rm H}$, versus $n_{\rm H}$. The lower
panel shows the {\it absolute} ion densities 
$n_{\rm ion}$ versus $n_{\rm H}$.  Because we have assumed
optically thin conditions, and fixed temperature,
both the relative and absolute
ion densities scale linearly with the metallicity.
 
The ``optimal'' hydrogen density
required to produce a particular ion depends on
the ionization parameter
\begin{equation}
U\equiv \frac{4\pi J^*}{n_{\rm H}c}
\end{equation}
where the Lyc integral
\begin{equation}
4\pi J^* = \int d\Omega\int_{{\nu_0}}^\infty d\nu \frac{J_\nu}{h\nu} \ \ \ ,
\end{equation} 
and $c$ is the speed of light.
For our standard and maximal fields, 
$4\pi J^*$ equals $1.3\times 10^4$ and $2.2\times 10^4$
photons s$^{-1}$ cm$^{-2}$ respectively.
The gas becomes more highly ionized when $U$ is large.
Thus, in Figure \ref{abun} where the radiation field
is held constant,  the ionization state increases
with decreasing gas density, as $U$ varies from $4.3\times10^{-7}$
to $13.6$.
The order of the various peaks in Figure \ref{abun}a,
is consistent with the photon energies required to produce the ions,
\ion{O}{6}~(113.90eV), \ion{N}{5}~(77.47eV), \ion{C}{4}~(47.89eV), 
\ion{Si}{4}~(33.49eV), \ion{S}{3}~(23.33eV), and \ion{Si}{3}~(16.35eV).
Figure \ref{abun}a shows that for our standard field
the {\it fractional} abundances
of \ion{O}{6}, \ion{C}{4}, and \ion{Si}{3} are maximized at hydrogen gas
densities of $3\times10^{-6}$, $3\times10^{-5}$, and 
$6\times10^{-4}$~cm$^{-3}$ respectively. For the maximal field
the optimal densities are a factor $\sim 2$ larger.

The absolute ion density is also an important quantity. 
For a given length scale
the ion column densities are largest when
the ion densities
approach the maximal values indicated in Fig.~2b.
In our mini-halo models, for example, the path-lengths
are set by the cloud sizes 
and the ion columns are largest along lines-of-sight
that intersect the locations where the ion densities are largest
(see equation[\ref{e:high}] in \S 4).
It is evident from Fig.~2 that the locations
of the density maxima do not necessarily coincide with the 
fractional abundances maxima. The behavior is governed by how
rapidly the ionization efficiency varies with density.
For example, the \ion{O}{6} and \ion{C}{4}
densities peak close to the locations of the fractional abundances maxima.
In contrast, the \ion{Si}{3} density continues to grow as $n_{\rm H}$ is
increased above the density of
$6\times10^{-4}$ cm$^{-3}$ at which the \ion{Si}{3} 
fractional abundance is maximized. 
The maximum \ion{C}{4} density occurs at $n_{\rm H}=1\times 10^{-4}$ cm$^{-3}$.
For $T = 10^4$~K this corresponds to a
gas pressure $P/k \sim 2$~cm$^{-3}$~K
\footnote{ 
For our assumed
helium abundance $P=2.17n_{\rm H}kT$ for fully ionized gas.}.
At this pressure and density the
expected \ion{C}{4} density in the metagalactic field is
$\sim7\times10^{-10}$~cm$^{-3}$, for a metallicity of 0.1.

Figure \ref{abun} shows
the dependence of the ionization state versus hydrogen density
for an unattenuated field.
In the mini-halo models we discuss next,
the results displayed in Figure 2 are
directly applicable to the outer WIM layers where the gas
is close to the optically thin limit.
They become less accurate with increasing
cloud depth, as the radiation field is absorbed
and the various ionization rates are selectively diminished.


\section{Mini-halo models}

Our goal is to determine the metal column density distributions
and UV absorption line properties of gas clouds 
in low-mass ($10^8 - 2\times 10^9$ $M_\odot$) mini-halos that are photoionized
by the present-day metagalactic field. In particular,
we wish to  find conditions that yield 
metal ion column densities comparable to those observed 
in the high-velocity \ion{C}{4} absorbers. 
As a guide to these observations, we focus on the
results presented by Sembach et al.~(1999; 2000) and 
Collins et al.~(2004) for the
\ion{C}{4} HVCs towards the extragalactic
sources Mrk~509 and PKS~2155-304. 
In Table \ref{sembach} we list the low-ionization 
\ion{C}{2},  \ion{N}{1}, \ion{Si}{2}, \ion{S}{2}, and high-ionization
\ion{C}{4}, \ion{N}{5},
\ion{Si}{3}, \ion{Si}{4}, and \ion{S}{3} column densities
inferred by Sembach et al.~(1999) 
for each of two high-velocity components observed towards each line of sight.
Sembach et al. (2000; 2003) detected
\ion{O}{6} absorption along these lines-of-sight,
and we include the inferred \ion{O}{6} columns in Table \ref{sembach}.  
More recently, Collins et al.~(2004) have presented new data on the
Mrk~509 and PKS~2155-304 absorbers, including observations obtained
with HST STIS. We summarize the Collins et al. data in Table
\ref{collins}.

We present two classes of illustrative models. 
First, we analyze the photoionization structures of
the low-mass ($\sim 10^8$ $M_{\odot}$), 
{\it pressure-confined}, mini-halo clouds
that were presented by SMW as representative models
for the \ion{H}{1} CHVCs. We recall that these 
two models were constructed by SMW to match the observed CHVC \ion{H}{1}
thermal phase properties, assuming ``circumgalactic'' or ``extragalactic''
distances of 150 or 750 kpc for these objects.
The characteristic mass-scale for the CHVC mini-halos is
set by the number of such objects ($\gtrsim 200$) in
comparison with theoretical simulations
that predict a steeply rising mass-spectrum towards low mass objects
in the dark-matter substructure
(Moore et al.~1999; Klypin et al.~1999; SMW). 
As we will show, the assumed external pressures 
($P_{\rm HIM}/k \sim 50$ cm$^{-3}$ K)
in these models severely limit the abundances of the high ions
produced by photoionization in the WIM envelopes.
Second, we consider {\it gravitationally-confined}
clouds in
more massive ($2\times 10^9$ $M_\odot$) ``dwarf-galaxy'' scale
halos, in which enhanced 
gravitational binding of the gas allows for lower external
confining pressures and hence increased ionization in the WIM.

In Table \ref{models} we list the basic input parameters for the
various models. These include, the halo virial 
mass $M_{\rm vir}$, concentration parameter
$x_{\rm vir}$, scale radius and velocity, $r_s$ and $v_s$,
radiation field (standard or maximal), 
total gas mass $M_{\rm WM}$  
(ionized plus neutral), metallicity, and external HIM pressure $P_{\rm HIM}$.

The parameters we have chosen for the two
pressure-confined systems are identical to those set by SMW
for the CHVCs as circumgalactic or extragalactic objects.
We refer to these as models A and B. 
The circumgalactic CHVC (model A) is a median ($\sigma=0$)
Burkert halo with 
$M_{\rm vir} = 1.0\times 10^8$~$M_\odot$ and $x_{\rm vir}=32.5$,
containing a total gas mass $M_{\rm WM}=1.1\times 10^6$~$M_\odot$ with
0.1~solar metallicity, exposed to the standard metagalactic field, and
subjected to an external pressure $P_{\rm HIM}/k= 50$~cm$^{-3}$~K.
SMW invoked a hot $2\times 10^6$~K
Galactic corona as the source of the required HIM pressure at 150 kpc.
The extragalactic CHVC (model B) is a $-4\sigma$ underconcentrated
$3.2\times 10^8$~$M_\odot$ Burkert halo with $x_{\rm vir}=8.2$,
$M_{\rm WM}=7.1\times 10^7$~$M_\odot$, again with
0.1 solar metallicity and exposed to the standard field,
with an assumed IGM bounding pressure $P_{\rm HIM}=20$~cm$^{-3}$~K.
Both models were constructed such that
the resulting peak \ion{H}{1} column densities 
are within the observed range of 
$2\times 10^{19}$ to $2\times 10^{20}$~cm$^{-2}$ for the CHVCs. 
 The relatively high external pressures in these models ensures that, 
(a) the gas remains bound 
\footnote{We define
particles as ``bound'' if
their gravitational potential energies exceed their mean
thermal energies.}
to the halos at all cloud radii, and that (b) the gas is compressed
to densities that allow the formation of neutral hydrogen
cores (which are then detectable as 21 cm sources).
For such low-mass halos, the gas becomes ionized and then unbound
as $P_{\rm HIM}$ is reduced.   

Figures \ref{circum} and \ref{extra} show the hydrogen 
and metal ion density
distributions and projected column densities
for our pressure-confined models.
In model A, the neutral WNM gas core extends to
a radius $r_{\rm WNM} = 0.67$ kpc, and the WIM shielding envelope
extends to an outer cloud radius $r_{\rm W/H} = 1.3$ kpc.
In model B, $r_{\rm WNM}=3.4$ kpc, and $r_{\rm W/H}=7.2$ kpc.
The low-ions, \ion{C}{2}, \ion{N}{1}, \ion{O}{1},
\ion{Si}{2}, and \ion{S}{2}, are the dominant ionization stages in 
the WNM. The high-ions, 
\ion{C}{4}, \ion{N}{5}, \ion{O}{6}, \ion{Si}{3},
\ion{Si}{4}, and \ion{S}{3}, are produced 
primarily in the WIM envelopes.
These are general properties of all of the models in our
parameter space. In the WNM, carbon, silicon, and sulfur are 
maintained in singly ionized form, while
nitrogen and oxygen, with ionization potentials only
slightly greater than 13.6 eV,
remain neutral. In the WNM, the densities of the low-ions 
are simply proportional to the
total hydrogen gas density, and are equal to the total metal abundances. 
The column densities of the low-ions are 
proportional to the line-of-sight \ion{H}{1} columns through the WNM. 
Observations of low-ions
along lines-of-sight through such \ion{H}{1} WNM cores
may therefore provide direct measures of the metallicity
when photoionization is the dominant ionization mechanism in the cores.

For the Burkert halos we are considering, the 
low-ionization metal columns at
$r=0$ may be approximated by
\begin{equation}
N_{\rm low}
=x_{\rm metal}N_{\rm HI,0}
\simeq \sqrt{\pi}x_{\rm metal}n_{\rm H,0}r_{\rm gas}
\end{equation}
where $N_{\rm HI,0}$ is the central (peak) \ion{H}{1} column,
$x_{\rm metal}$ is the metal abundance,
$n_{\rm H,0}$ is the hydrogen gas density at the cloud center,
and $r_{\rm gas}$ is the $1/e$ gas scale height for the neutral gas
\footnote{The relation 
$N_{\rm HI,0}=\sqrt{\pi}n_{\rm H,0}r_{\rm gas}$ for Burkert halos is
exact in the gravitationally-confined 
limit which obtains when $v_s/c_g \gtrsim 2$,
where $c_g$ is the sound speed. In this limit 
$r_{\rm gas}=0.44\times 10^{-0.175\sigma}c_{g6}M_{\rm vir,8}^{0.100}$ kpc,
where $c_{g6}$ is the sound speed in units of 10 km s$^{-1}$,
and $M_{\rm vir,8}$ is the virial mass in units of $10^8$ M$_\odot$.
These expressions are accurate
to within a factor of $\sim 2$ for $v_s/c_g = 1.5$ as
is the case for our CHVC models (see SMW for a discussion of
these analytic expressions).}.
For example, in Figure \ref{circum}, $N_{\rm HI,0}=5\times10^{19}$~cm$^{-2}$ 
and $N_{\rm OI,0}=3\times10^{15}$~cm$^{-2}$, consistent with our assumed 
oxygen abundance
of $5.4\times 10^{-5}$.  The column densities of
the low-ions in models A and B are comparable
because the \ion{H}{1} columns in these models
are approximately the same by construction. 

In both models A and B, \ion{C}{2}, \ion{Si}{2}, and \ion{S}{2} remain
the dominant stages in the WIM. However, \ion{N}{1} and \ion{O}{1} 
decrease sharply beyond the ionization fronts and into the
WIM where these neutral stages are efficiently removed
by photoionization.
The densities of the high-ions produced in the WIM
are controlled by several competing effects. 
First, the ionization state of the gas is largest near the cloud edges, where
the ionization parameter is largest, i.e., where 
the gas density is lowest and where the radiation field is least attenuated.
But second, the absolute ion densities may be maximized
in inner regions where the overall gas densities are larger,
or where the ionization parameter is optimized for the production
of the particular ion.
Third, the absorption of the ionizing radiation,
especially near the ionization front, modifies the ionization structure.
Figure \ref{abun}b may be used as a guide to the
gas densities and cloud radii at which the density of a
particular high-ion may be expected to be most
efficiently produced, neglecting optical depth effects.
For example, in models A and B
the \ion{C}{4}, \ion{Si}{3}, and \ion{Si}{4} densities
are largest at the cloud edge. In model A, where
$n_{\rm H}=2.3\times 10^{-3}$~cm$^{-3}$ at $r_{\rm W/H}$,
the densities of these ions equal $8.8\times10^{-11}$, 
$3.4\times10^{-9}$, and $1.0\times10^{-10}$~cm$^{-3}$ 
respectively, as expected from Figure \ref{abun}b.
In model B the column densities of the
high-ions are $\sim 10$ times larger,
because of the slightly lower
bounding pressure and larger ionization parameter at the cloud edge,
but more importantly because of the significantly larger cloud size.
Negligible amounts of \ion{N}{5} and \ion{O}{6} are produced in these models.

To order-of-magnitude, the maximal column density
of a high-ion may be expressed as
\begin{equation}
\label{e:high}
N_{\rm high} \simeq 2{\rm max}[n_{\rm high}]r_{\rm W/H}
\end{equation}
where ${\rm max}[n_{\rm high}]$ is the maximum ion density
in the cloud, and where we adopt the cloud diameter $2r_{\rm W/H}$
as a characteristic length scale for the WIM. 
For example, for model A, 
where $n_{\rm CIV,max}=8.8\times10^{-11}$ cm$^{-3}$ and $r_{\rm W/H}=1.3$ kpc,
equation (\ref{e:high}) gives a peak \ion{C}{4} column
density of $7.1\times10^{11}$~cm$^{-2}$, roughly consistent with the maximal
column of $2.8\times10^{11}$~cm$^{-2}$
occurring at a projected radius of 1.1 kpc (see Figure \ref{circum}b).

For a given bounding pressure, the cloud size $2r_{W/H}$ is 
proportional to the $1/e$ gas scale height for the ionized gas.
It therefore follows from equations (8) and (9) that
\begin{equation}
\label{low-high-ratio}
\frac{N_{\rm high}}{N_{\rm low}} \propto 
\frac{{\rm max}(n_{\rm high})}{n_{\rm H,0}} \ \ \ .
\end{equation}
The density max$(n_{\rm high})$ is fixed by the external
pressure and radiation field. However,
the central density $n_{\rm H,0}$ depends
on the halo potential and becomes small
for underconcentrated halos. Thus, $N_{\rm high}/N_{\rm low}$
may be used as a dynamical probe of the halo concentration
parameter. For example, the \ion{C}{4}/\ion{C}{2} column
density ratio increases from $1.1\times10^{-4}$ to 
$9.6\times10^{-4}$ from model A to B.
The increase in the \ion{C}{4}/\ion{C}{2} ratio is primarily due
to the increase in cloud size, from model A to B,
as the concentration parameter decreases from 32.5 to 8.15.

In Table \ref{ceres} we list the \ion{H}{1} and 
metal absorption-line column densities
along lines-of-sight that are coincident with the maximal \ion{C}{4}
columns in models A and B.  
Additional output parameters are also listed.
It is immediately evident that
the computed column densities are largely inconsistent
with the high-velocity absorbers towards Mrk~509 and PKS~2155-304.
The predicted \ion{C}{4} and \ion{Si}{4} columns are several orders
of magnitude below the observed values, and the
\ion{H}{1} and \ion{C}{2} columns are too large.
The primary difficulty is that in our pressure-confined models
the ionization-parameter is too low.

It is clear that to increase the column densities of
species such as \ion{C}{4} and \ion{Si}{4}, via photoionization,
to observed values exceeding a few $10^{13}$ cm$^{-2}$,
the bounding pressure must be reduced significantly.
However, if this is done for our low-mass CHVC mini-halos,
the gas becomes unbound at the larger radii where such species form with
greater efficiency
\footnote{This follows directly from the fact that in
models A \& B the gas at $r_{\rm W/H}$ is just marginally
bound for the assumed bounding pressures of 50 and 20~cm$^{-3}$~K.}.
We therefore consider more massive, ``dwarf-galaxy'' scale
halos, in which the clouds are primarily gravitationally confined.
In these models the gas
remains bound even when the external pressure 
is reduced to very low values.  

In \S 3 we found that 
the optimal pressure for the production of \ion{C}{4}
in the standard metagalactic field is $P/k \approx 2$~cm$^{-3}$ K,
and that for 0.1 solar metallicity the \ion{C}{4} density then equals
$\sim 7\times 10^{-10}$~cm$^{-3}$. 
For these conditions, a column
density of $\sim 5\times 10^{13}$~cm$^{-2}$ requires a
path-length of 25~kpc. 
This length scale is
much larger than the optical diameters of dwarf galaxies,
or of the larger \ion{H}{1} halos surrounding them 
(Mateo 1998; Young et al.~2003).
Very tenuous and extended WIM envelopes
might exist around the dwarfs, especially if the halos
are very underconcentrated.
Alternatively, smaller path-lengths are required if
the metallicity and photoionization rates are both
enhanced by factors of a few. We consider these
possibilities below.

We focus on a series of models, labeled
C through J, with input parameters as listed in Table \ref{models}. For most
of these we adopt median ($\sigma=0$) Burkert potentials, 
and assume representative
halo masses $M_{\rm vir}=2\times 10^9$~$M_\odot$ 
with halo scale velocities $v_s=30.0$ km s$^{-1}$.
Two exceptions are model H, which is a $-5\sigma$ underconcentrated
Burkert halo, and model I, in which we adopt an NFW profile. 
In all models (except J) we choose a total gas mass such that
the resulting
peak \ion{H}{1} columns are in the range of $1-2\times 10^{21}$~cm$^{-2}$
as observed in the Local Group dwarf galaxies Leo A and Sag DIG
(Young \& Lo 1996; 1997).
In model J we reduce the total gas mass such 
the cloud consists of WIM only, without a neutral core.

The parameters for model C are identical to those adopted by
SMW in their dwarf galaxy model, and include
a small contribution of stars to the total gravitational potential.
In our models
we vary the bounding pressures from 1.0 to 0.01 cm$^{-3}$ K,
consider photoionization by the ``standard'' or ``maximal''
radiation fields, and increase the metallicity from 0.1 to 0.3.
The resulting
photoionization structures are displayed in Figures \ref{dwarf} through 
\ref{bigImax0.3}.
In Table \ref{dwarfres} we list output data,
including the \ion{H}{1} and metal column densities along the
representative lines-of-sight 
for which the \ion{C}{4} columns are largest.
 
In models C, D and E, displayed in Figures \ref{dwarf}, \ref{bigdwarf} 
and \ref{hugedwarf},
we assume photoionization by 
the standard field, and a metallicity of 0.1.
In this sequence we reduce the bounding pressure from
1.0, to 0.1 to 0.01~cm$^{-3}$~K 
respectively. In these models most of the gas is WNM,
with $M_{\rm WNM} \cong 1.2\times 10^7$~$M_\odot$, and
the ionization fronts occur at fixed radii $r_{\rm WNM}=1.2$ kpc.
The outer edges of the WIM envelopes occur at $r_{\rm W/H}$
equal to 4.2, 7.7, and 13.6~kpc. Because the
gas densities decrease rapidly with radius the WIM masses are essentially
independent of the bounding pressures, and
$M_{\rm WIM} \cong 4\times10^6$ $M_\odot$.

A plausible lower-bound to the IGM pressure in the Local Group may be
set by adopting the mean cosmic baryon density 
$n_b=2.5\times 10^{-7}$ cm$^{-3}$ as a minimum density
and assuming a
virial temperature $T\approx 4.3\times 10^5$ K corresponding
to the Local Group 
velocity dispersion of 60 km s$^{-1}$ (Sandage 1986).
Thus, at large distances from the Galaxy and M31, where
the overdensities may be low, IGM pressures as low as 0.2 cm$^{-3}$ K
may prevail. Recent
{\it Chandra} and {\it XMM-Newton} X-ray absorption-line
observations of highly ionized \ion{O}{7} and \ion{O}{8}
imply pressures in the range 0.5-3.5 cm$^{-3}$ K
for hot gas in the Local Group (Nicastro et al.~2002; see also
Rasmussen, Kahn \& Paerls 2003).
We note further that for $10^4$ K gas, and a pressure of 0.1 cm$^{-3}$ K
the gas recombination time is $\sim 2.6\times10^{10}$ yr, so that 
ionization equilibrium for the high-ions
can be reached within a Hubble time at
such or larger pressures. 

As expected, 
in the WNM cores of our dwarf galaxy models
the column densities of the low-ions are proportional to
$N_{\rm HI}$. 
The \ion{N}{1} and \ion{O}{1} densities
again fall sharply beyond the ionization fronts, but in contrast to 
our CHVC models,
the outer parts of the WIM envelopes become highly ionized.
In particular, in our dwarf galaxy models
\ion{C}{2}, \ion{Si}{2}, and \ion{S}{2}, are no longer the dominant
ionization stages in the outer parts of the WIM.
In models C, D, and E, the \ion{C}{4} density reaches 
maximal values of $7.2\times10^{-10}$ at $r\approx 3.2$~kpc where
$P/k\approx 3.2$~cm$^{-3}$~K . The
peak \ion{C}{4} column density increases 
as $P_{\rm HIM}$ is reduced from 1 to 0.1~cm$^{-3}$
because as the pressure is reduced the cloud size and projected
path-lengths increase. However, reducing the pressure further
to 0.01~cm$^{-3}$ does
not lead to a significant increase in the \ion{C}{4} 
column because of the rapidly
declining ion densities at these pressures.
In these models, 
peak \ion{C}{4} columns of $1-2\times 10^{13}$~cm$^{-2}$
occur at projected radii $\sim 2.6$ kpc from the cloud centers.
The column densities are
consistent with equation (\ref{e:high}) to order-of-magnitude.

As indicated by Figure 2, \ion{O}{6} is
maximized at $P/k=0.1$~cm$^{-3}$~K, or $n_{\rm H}=1\times10^{-5}$~cm$^{-3}$.
For a metallicity of 0.1, the optimal \ion{O}{6} density is 
$1\times10^{-10}$~cm$^{-3}$.
Thus, in models C and D, the peak \ion{O}{6} columns increase
from $2.9\times10^{11}$ to $2.7\times10^{12}$~cm$^{-2}$ as 
$P_{\rm HIM}$ is reduced from 1 to 0.1 cm$^{-3}$ K.
However, reducing the pressure further as in model E
does not lead to a significant increase in the \ion{O}{6} columns.

The predicted \ion{C}{4} and \ion{Si}{4} columns in this first
set of dwarf galaxy models are much larger than 
we found for the CHVCs,
but are about a factor of ten smaller
than those observed in the \ion{C}{4} HVCs.
However, increasing the radiation intensity and 
the metallicity can bring the predicted 
columns into harmony with the observations.
Thus, in model F we set $P_{\rm HIM}/k=1$~cm$^{-3}$, 
and keep the metallicity
at 0.1, but now adopt our
maximal radiation field as given by
equation (\ref{radm}). As shown by Figure \ref{maxspec},
the elevated photoionization rates
increase the resulting WIM metal columns by
factors of 2-3. In model G 
we increase the column densities further (see Figure \ref{big0.3max}) 
by lowering the bounding
pressure to 0.1 cm$^{-3}$ (thereby increasing the \ion{C}{4} pathlengths),
by increasing the metallicity to 0.3, and again using the maximal field.
In model G the peak \ion{C}{4} column density is 
$1.5\times 10^{14}$ cm$^{-2}$, comparable to or greater than 
the observed values in the \ion{C}{4} absorbers.


With the exception of \ion{O}{6}, model G provides
a satisfactory match to the relative and absolute metal column densities
in the -283 and -228 km s$^{-1}$ \ion{C}{4} 
absorbers towards Mrk~509, and the -256 km s$^{-1}$ absorber
towards PKS 2155-304. An interesting difficulty is encountered
for the -140 km s$^{-1}$ PKS 2155-304 absorber.
As an illustrative example, in Figure 9c we plot the ratios,
$N_{\rm model}/N_{\rm observed}$, of the columns computed
in model G and the
observed columns in the $-283$~km~s$^{-1}$ Mrk~509 absorber
(see Tables \ref{sembach} and \ref{collins}),
as functions of impact parameter in model G.
This plot reveals the range of model G sight-lines
that are consistent with the observed ion column densities,
including the upper and lower-limits.
In the shaded area in Figure 9c the model and measured columns 
are in agreement to within $\pm 0.5$ dex.
For each of the species, the arrows 
indicate the range of radii (between the arrowheads) where the 
model columns are within 0.5 dex of the measured values or limits.
A successful fit is achieved if all of the indicated ranges overlap.

Figure 9c shows that lines-of-sight
from $0.0$ to $3.9$~kpc from the cloud center
produce an adequate amount of 
\ion{C}{4}. No constraints are provided by 
\ion{N}{5} and \ion{S}{3},
for which the computed values are consistent with the 
observed upper-limits for all lines-of-sight.
The \ion{N}{1}, \ion{Si}{2}, and \ion{S}{2} upper-limits
are consistent with
impact parameters of at least $1.2$, $1.8$, and $1.0$ kpc respectively.
A strong constraint is provided by 
\ion{C}{2}, which restricts the projected radii to the range $2-2.6$~kpc.
The measured column densities of all other species, with the
exception of \ion{O}{6}, are consistent with the model for
$2-2.6$ kpc. The observed \ion{O}{6} column cannot be
reproduced for any line-of-sight.
We note that the line-of-sight for which \ion{C}{4}
is maximized occurs at $2.3$~kpc, within the range constrained by \ion{C}{2}.
For this radius,
the predicted \ion{C}{4}, \ion{Si}{3}, and \ion{Si}{4} columns 
are equal to
$1.5\times 10^{14}$, $3.3\times 10^{13}$, and $2.0\times 10^{13}$~cm$^{-2}$,
respectively,
compared to the observed columns, 
of $\sim 1.5\times 10^{14}$~cm$^{-2}$
for \ion{C}{4}, $\sim 2\times 10^{13}$~cm$^{-2}$ for \ion{Si}{3},
and $\sim 3\times 10^{13}$~cm$^{-2}$ for \ion{Si}{4}.
At 2.3 kpc the predicted \ion{H}{1} and
\ion{C}{2} columns of $1.8\times 10^{16}$ and
$6.5\times 10^{13}$~cm$^{-2}$ are consistent with
the observed \ion{H}{1} upper-limit and measured \ion{C}{2}
column of $4.3\times 10^{-2}$~cm$^{-2}$.  As already indicated,
at this radius, the remaining computed
columns are consistent
with all of the upper-limits. Of specific interest is \ion{N}{5}, 
for which our predicted
column of $7.1\times 10^{12}$~cm$^{-2}$ is close to
the observed upper-limit of $1.5\times 10^{13}$~cm$^{-2}$.

Similar fits may be found for the
$-228$~km~s$^{-1}$ Mrk~509 and 
$-256$~km~s$^{-1}$ PKS~2155-304 absorbers.
The range of radii in model G that are 
consistent with the data (excepting \ion{O}{6}) in these absorbers
are 3.1 to 3.4~kpc and 3.2 to 3.5~kpc, respectively. 
However, the $-140$~km~s$^{-1}$ PKS~2155-304 absorber
appears anomalous in that the
abundance ratio of low-ions to high-ions is
larger in this system compared to the other absorbers. 
In particular, sight lines with impact parameter between
$3.1$ to $5.0$~kpc consistent with the \ion{C}{4} columns are
underabundant in the low-ions \ion{C}{2}, \ion{Si}{2} and 
\ion{Si}{3} and also \ion{H}{1}, while more centrally confined 
sight lines, from $2.1$ to $2.5$~kpc are consistent with these 
low ions, but yield an excess of \ion{C}{4}.

We conclude that photoionized WIM envelopes around dwarf galaxies 
are natural sites for the production of
\ion{C}{4}, \ion{Si}{3}, and \ion{Si}{4} ions,
and the simultaneous reduction of low-ions such
as \ion{C}{2}, \ion{Si}{2}, and \ion{S}{2}.
Ion column densities ranging from
$10^{13}$ to $10^{14}$~cm$^{-2}$, comparable to
those in the \ion{C}{4}
HVCs, are possible, provided the IGM pressure is low 
with $P/k \lesssim 1$~cm$^{-3}$ K as found by Sembach et al.~(1999). 
If median $\Lambda$CDM Burkert halos are good representations
of dwarf galaxy dark-matter halos,
the implied gas path-lengths require 
metallicities $\gtrsim 0.3$ to produce such large metal
columns in individual dwarf galaxies.
However, photoionization by metagalactic radiation is
not an efficient source of \ion{N}{5} or
\ion{O}{6} in such WIM envelopes.

Model H is a $-5\sigma$ Burkert halo, and illustrates
the effect of reducing the halo concentration
parameter. In making this reduction we keep 
$v_s$ fixed at 30.0 km s$^{-1}$. The associated halo mass
is then $M_{\rm vir}=6.2\times 10^9$ $M_\odot$.
In Figure \ref{-5sig} we display results for the standard field,
$P_{\rm HIM}/k =1$ cm$^{-3}$ K, and a metallicity of 0.1.
We choose a gas mass $M_{\rm WM}=1.3\times 10^9$ that yields
a peak \ion{H}{1} column of $1.4\times 10^{21}$ cm$^{-2}$.
The column densities of the low-ions in the WNM
core are therefore unchanged with respect to the 
corresponding median halo model C. However,
the cloud size in model H is significantly larger with
$r_{\rm W/H}=25$ kpc. The column densities of the high-ions
are therefore enhanced. For example the peak \ion{C}{4}
column for this model is $7.5\times 10^{13}$ cm$^{-2}$.
In moving from model C to H
the \ion{C}{4}/\ion{C}{2} column density ratio increases from
$2.4\times10^{-4}$ to $1.3\times10^{-3}$, consistent with 
equation (\ref{low-high-ratio}). 

In model I, displayed in Figure \ref{-4sigNFW},
we consider a dwarf galaxy model using a NFW profile.
We adopt a $-4\sigma$ underconcentrated halo to yield a neutral
gas scale height similar to those in our median Burkert models
which match the observed HI gas distributions in Leo A and Sag DIG (SMW).
We again fix $v_s=30$~km~s$^{-1}$. The virial mass is then
$4.9\times10^9$~$M_\odot$. For a direct comparison with model G
we show results for the maximal field, $P_{\rm HIM}=0.1$~cm$^{-3}$~K, and a
metallicity of 0.3.  A total gas mass of $3.1\times 10^7$~$M_\odot$
is required to produce an \ion{H}{1} column of $1.4\times10^{21}$~cm$^{-2}$.
The projected neutral gas 1/e gas scale height is 0.4 kpc. 
In our NFW model 
the cloud radius $r_{\rm W/H}=12.4$~kpc (as opposed to 7.4 kpc
in model G). The larger size results in part from the fact
that the gas distributions in NFW potentials are
exponentials and decrease less rapidly than in
Burkert potentials where the gas distributions are Gaussians (SMW).
The larger cloud radius leads to correspondingly larger column 
densities for the high-ions. The peak \ion{C}{4}
column in this model is $2.6\times 10^{14}$ cm$^{-2}$.  

In the dwarf galaxy models we have discussed so far, the WIM components
exist as envelopes around the WNM cores. Furthermore, the large opacities
associated with the peak \ion{H}{1} columns of $1.4\times 10^{21}$ cm$^{-2}$
in these models imply that the external metagalactic field cannot
alone provide the gas heating for the WNM. As found by SMW, 
for median Burkert halos with $M_{\rm vir}=2\times 10^9$ $M_\odot$
a maximum gas mass of $3\times 10^6$ $M_\odot$ can be maintained
at $\sim 10^4$ K by the present-day metagalactic field. 
Internal heat sources, such as stars,
are required to maintain larger WNM masses in dwarf
galaxies.  However, the properties of the WIM envelopes, which may be
fully ionized and heated by the external field, do not depend
significantly on the presence of the WNM cores. In particular,
``fully ionized dwarf galaxies'' could exist,
in which the dark-matter halos never contained 
sufficient gas mass for the formation
of neutral cores and stars, but which contain photoionized gas
detectable via UV metal absorption lines.  

To demonstrate this, we conclude with model J, which represents such
a fully ionized dwarf galaxy. The parameters are identical to
model G, except that we reduce the gas mass to 
$M_{\rm WM}=1\times 10^6$ cm$^{-2}$.  The resulting peak \ion{H}{1}
column density is now only $5\times 10^{17}$ cm$^{-2}$, so
the cloud remains optically thin and fully ionized.
The gas cloud is entirely WIM. The cloud radius $r_{\rm W/H}=4.7$ kpc
is smaller than in model G because the WNM core has been removed.
As is indicated by Figure \ref{bigImax0.3} and Table \ref{dwarfres}
the resulting metal columns are reduced, but only by about 30\%.


\section{Discussion and Summary}

In this paper we study the distributions and relative abundances 
of metal ions in gas clouds that are confined in low-mass dark-matter
mini-halos in the extended Galactic halo or Local Group, and
which are photoionized by the present-day metagalactic radiation field. 
We consider spherical and hydrostatic clouds, with gas
temperatures $\sim 10^4$ K, which are also subjected to
external bounding pressures provided by a hot medium.
We compute the distributions of low-ions,
\ion{C}{2}, \ion{N}{1}, \ion{O}{1}, \ion{Si}{2}, and \ion{S}{2},
produced primarily in the optically thick WNM cores,
and high-ions,
\ion{C}{4}, \ion{N}{5}, \ion{O}{6}, \ion{Si}{3}, \ion{Si}{4}, and \ion{S}{3},
produced in the outer WIM shielding envelopes of the clouds.
We study the ionization properties of the
compact \ion{H}{1} high-velocity clouds, as photoionized pressure-confined
gas clouds in $\sim 10^8$ M$_\odot$ mini-halos. We then focus
on the metal ion distributions in more massive 
$\sim 2\times 10^9$ M$_\odot$ halos at the dwarf-galaxy scale. 
In particular, we
determine conditions for the formation of highly-ionized
components, with ionization states similar to the high-velocity
\ion{C}{4} absorbers observed by Sembach et al.~(1999)
and Collins et al.~(2004) towards Mrk~509 and PKS~2155-304. 

First, we address the question of whether the
\ion{C}{4} HVCs could be the ionized counterparts or envelopes of
pressure-confined \ion{H}{1} CHVCs, 
which are photoionized by metagalactic radiation, where
external bounding pressures $P/k \sim 50$~cm$^{-3}$~K
are required to bind, compress, and neutralize the \ion{H}{1} clouds.
We find that in such objects the ionization parameter 
is too low to allow efficient production of \ion{C}{4} and other
high-ions.
The dominant ionization stages in these objects remain 
\ion{C}{2}, \ion{N}{1}, \ion{O}{1}, \ion{Si}{2}, and \ion{S}{2}.
If the CHVCs are pressure-confined circumgalactic objects (as favored
by SMW) then significant columns of high-ions require alternative
ionization mechanisms, as suggested by Fox et al.~(2003)
for the relatively nearby Complex C.
One attractive possibility to be considered is the role
of non-equilibrium ionization in the conductive interfaces 
that must exist between
the WIM envelopes of mini-halo clouds and surrounding hot gas.
Our computations, and the recent \ion{O}{6} detections, confirm
the conclusions of Sembach et al.~(2000; 2003) that 
additional processes are likely affecting the WIM ionization properties
in the Mrk~509 and PKS~2155-304 absorbers.
In the WNM cores,
the column densities of the low-ions and atoms are predicted to be equal to
the metal abundance times the \ion{H}{1} column along the line-of-sight.
UV absorption-line observations of such low-ions
can therefore provide useful and direct measures of the CHVC metallicities,
if photoionization dominates in the cores.

Second, we study the properties of more extended and tenuous photoionized 
WIM envelopes around more massive, dwarf galaxy scale, mini-halos.
In such systems, gravitational confinement of the gas is more effective,
and the bounding pressures can be taken to much lower values.
The ionization parameters in the WIM may therefore be increased
to values that yield large abundances of 
\ion{C}{4}, \ion{S}{3}, \ion{Si}{4}, and \ion{Si}{3},
and simultaneously,
reduced abundances of low-ions.  We find that
ionization states comparable to those in the \ion{C}{4} HVCs
are produced in the WIM envelopes of dwarf galaxies
for bounding pressures $P/k \lesssim 1$~cm$^{-3}$~K.
An important exception is \ion{O}{6}, which is inefficiently produced by
photoionization.

In our models, 
the predicted metal column densities are proportional to
the cloud sizes and associated path-lengths. 
For a given gas velocity dispersion, the cloud
size-scale is fixed by the dark-matter gravitational potential.
For $\Lambda$CDM Burkert halos with median dark-matter concentration
parameters, we find that the
WIM envelopes extend out to 5-10~kpc from the cloud centers,
depending on the value of the bounding pressure.
\ion{C}{4} columns of up to $\sim 10^{14}$~cm$^{-2}$, as observed by
Sembach et al.~(1999), would then require metallicities of
at least 0.3 times solar for absorption occurring in a single WIM
envelope.  We conclude that the WIM envelopes of dwarf galaxies
are natural sites for the formation of 
\ion{C}{4}, \ion{S}{3}, \ion{Si}{4}, and \ion{Si}{3}
via metagalactic photoionization provided $P/k \lesssim 1$~cm$^{-3}$~K
in the surrounding IGM.
This is consistent with the finding of Sembach et al. (1999) that such
low pressures favor the production of \ion{C}{4}.
Such low pressures appear to be
indicated by recent X-ray absorption-line
observations of very highly-ionized \ion{O}{7} and \ion{O}{8} that may be
probing the hot IGM medium in the Local Group
(Nicastro et al.~2002; Rasmussen, Kahn \& Paerels 2003).

We also present results for fully ionized dwarf galaxies,
in which the gas mass confined to the mini-halos
remains optically thin.
In such objects WNM cores do not exist and the
\ion{H}{1} column densities are correspondingly low.
Such objects are also expected to be starless given that
photoionization was more effective at earlier epochs, at least
since reionization (Efstathiou 1992; Kepner, Babul \& Spergel 1997;
Barkana \& Loeb 1999).  Our computations show that such
``dark galaxies'' could be detectable as UV metal-line absorbers,
with ionization states similar to those observed in the
\ion{C}{4} absorbers. However, given the relatively high
metallicities required, the gas that was trapped in
such halos must have been previously enriched.

The origin, nature, and location of the \ion{C}{4} HVCs remain uncertain.
Sembach et al.~(1999) suggested that the absorbers towards
Mrk~509 and PKS~2155-304 may be related to the group of 
\ion{H}{1} HVCs observed in the 
``Galactic center negative'' (GCN) direction 
(Wakker \& van Woerden 1991).
Some of these GCN clouds have velocities close to the
\ion{C}{4} HVCs, and are located within 5$^\circ$ of them.
The GCN clouds may be tracing a system of low-surface
brightness galaxies (Mirabel \& Cohen 1979),
although an origin in the Magellanic Stream has also
been suggested (Wakker \& van Woerden 1991).
Our models suggest that a system of ionized dwarf galaxies,
perhaps associated with the GCN \ion{H}{1} clouds 
viewed as low-surface brightness galaxies,
could, with the exception of \ion{O}{6}, 
give rise to UV metal absorption features
similar to those observed in the \ion{C}{4} HVCs.

     
\section*{Acknowledgments}

We thank the US-Israel Binational Science Foundation 
(grant 2002317) for supporting our research.
We thank C.F.~McKee for discussions, and the anonymous referee
for helpful advice and suggestions.
We thank Gary Ferland for maintaining CLOUDY as a 
publicly available code.



\clearpage

\begin{figure}
\plotone{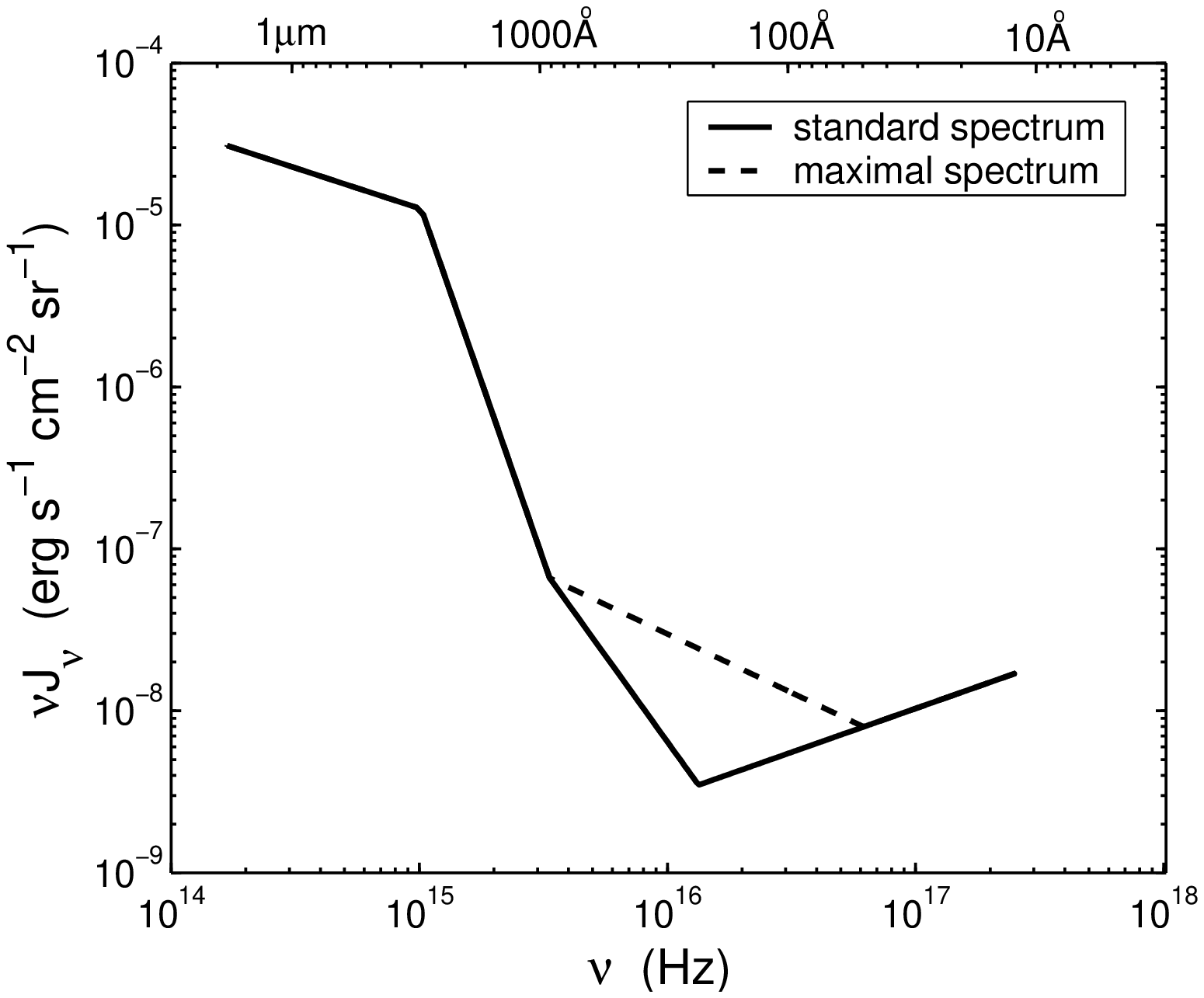}
\caption{The standard metagalactic radiation field (solid line) and 
the maximal field (dashed line).}
\label{radfield}
\end{figure}

\begin{figure}
\includegraphics{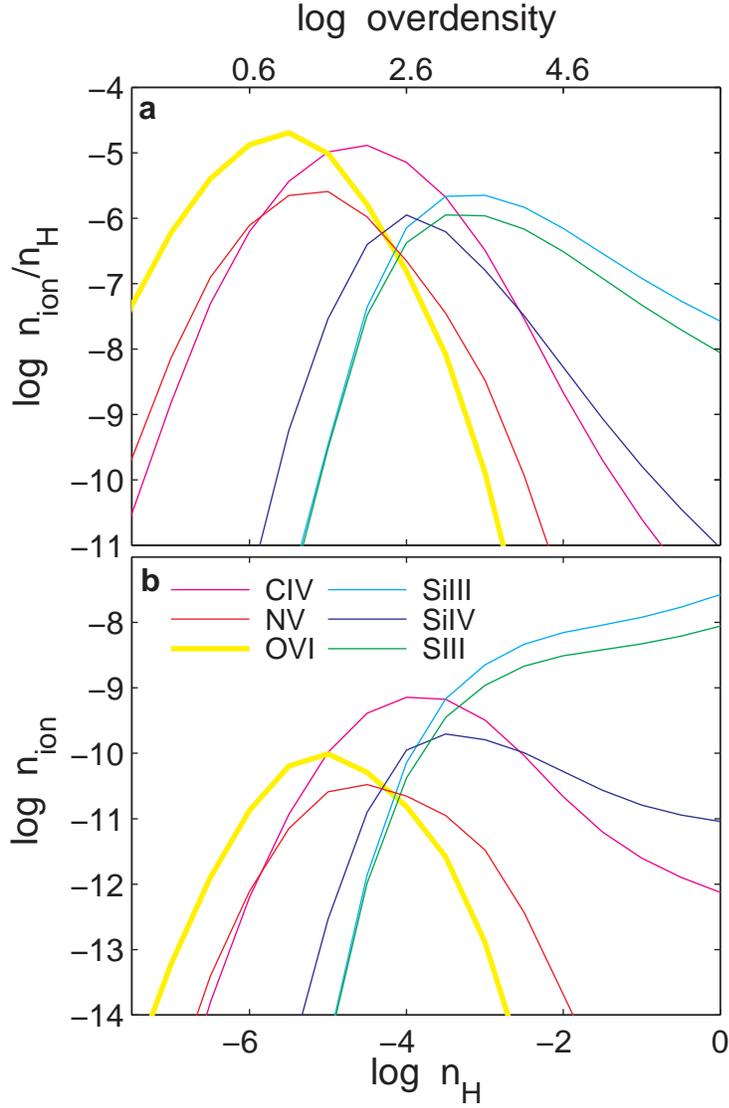}  
\caption{The abundances of 
high-ions as a function of the hydrogen density, 
for photoionization of $10^4$~K gas by the standard metagalactic field: 
panel (a) shows the abundances relative to hydrogen, 
and panel (b) shows the absolute abundances.
Color version is available in the electronic on-line edition.} 
\label{abun}
\end{figure}

\begin{figure}
\plotone{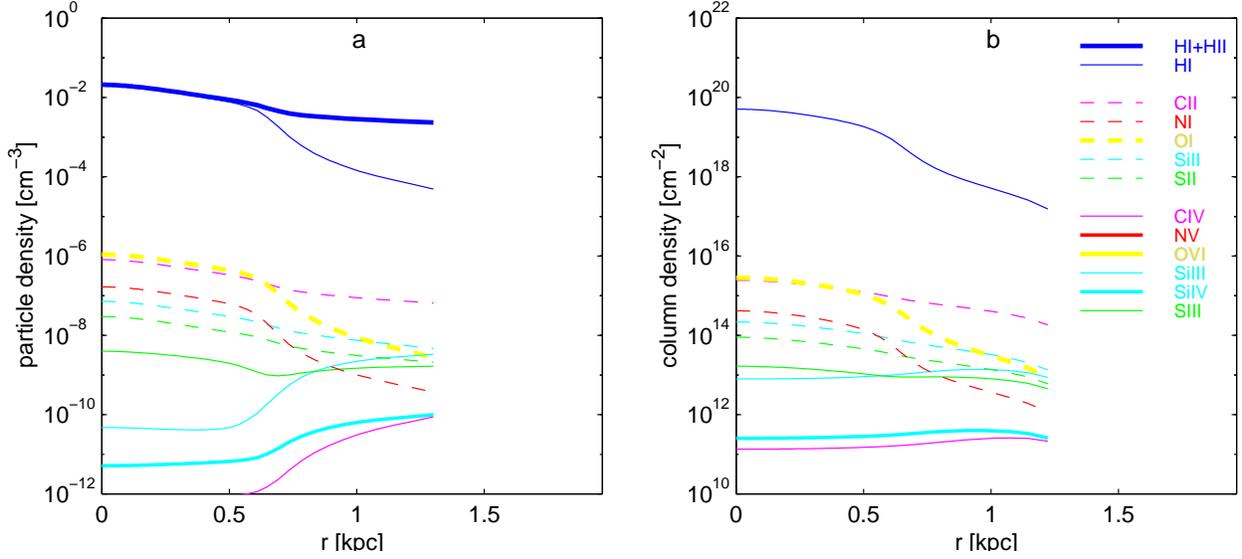}   
\caption{Model A:
$M_{vir} = 1\times10^8$~M$_{\odot}$ median halo, 
$M_{gas} = 1.1\times10^6$~M$_{\odot}$,
$0.1$~solar metallicity,
photoionization by the standard metagalactic field, bounding pressure 
$P_{\rm HIM}/k_B = 50$~cm$^{-3}$~K.
Panel (a) shows the volume densities as a function of the
distance from the cloud center, 
and panel (b) shows the projected column densities}
\label{circum}
\end{figure}

\begin{figure}
\plotone{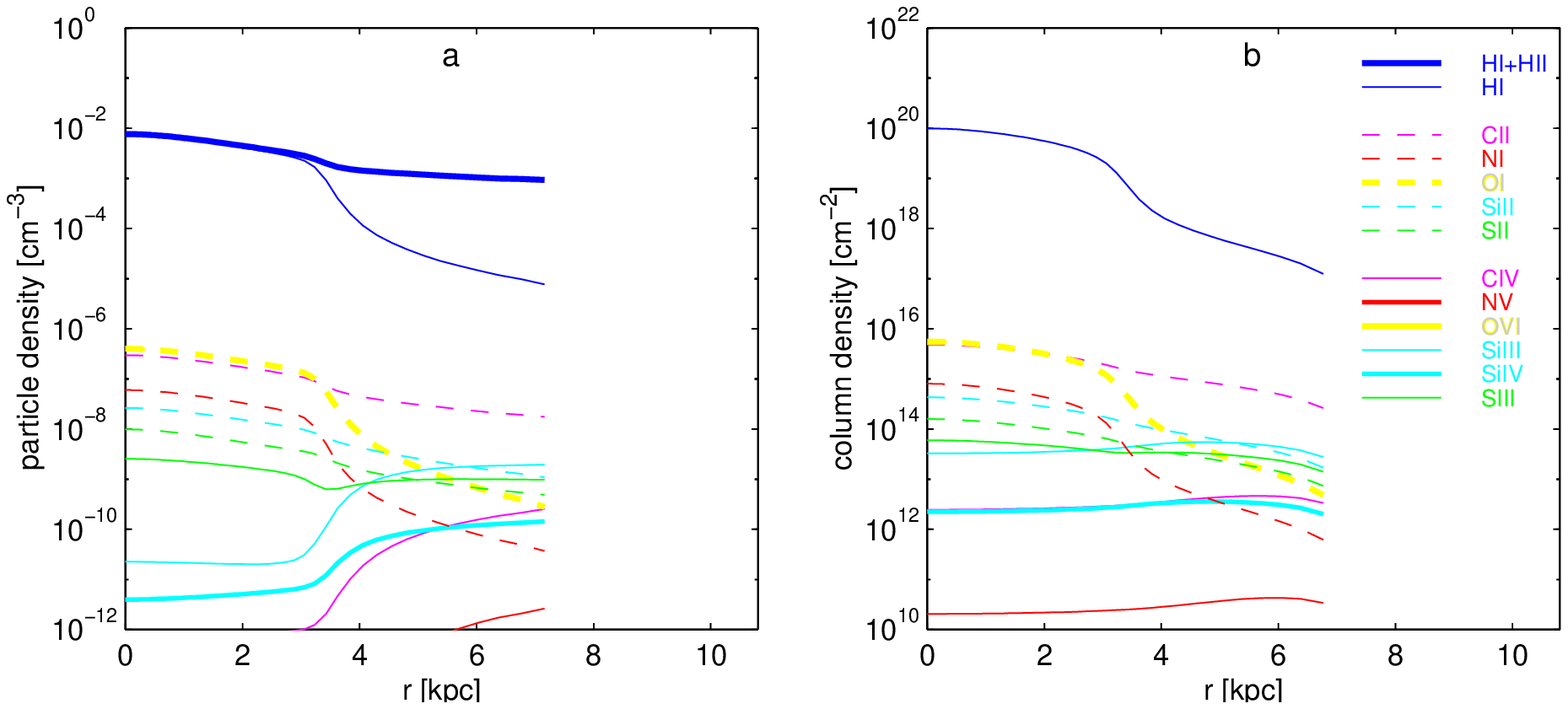}  
\caption{
Model B:
$M_{vir} = 3.2\times10^8$~M$_{\odot}$ $-4\sigma$ halo, 
$M_{gas} = 7.1\times10^7$~M$_{\odot}$,
$0.1$~solar metallicity,
standard radiation field, 
$P_{\rm HIM}/k_B = 20$~cm$^{-3}$~K.}
\label{extra}
\end{figure}

\begin{figure}
\plotone{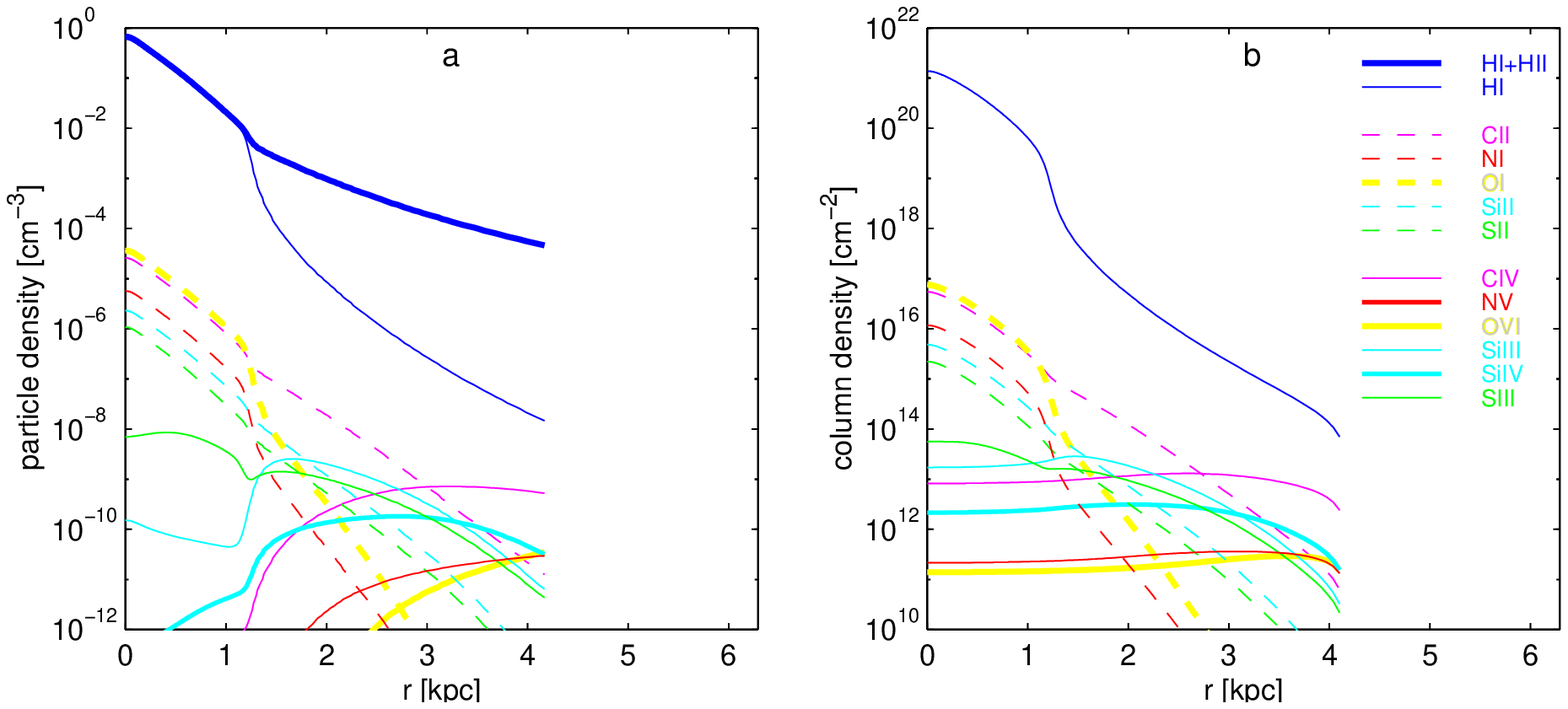}  
\caption{
Model C:
$M_{vir} = 2\times10^9$~M$_{\odot}$ median halo, 
$M_{gas} = 1.6\times10^7$~M$_{\odot}$,
$0.1$~solar metallicity,
standard radiation field, 
$P_{\rm HIM}/k_{B} = 1$~cm$^{-3}$~K.}
\label{dwarf}
\end{figure}

\begin{figure}
\plotone{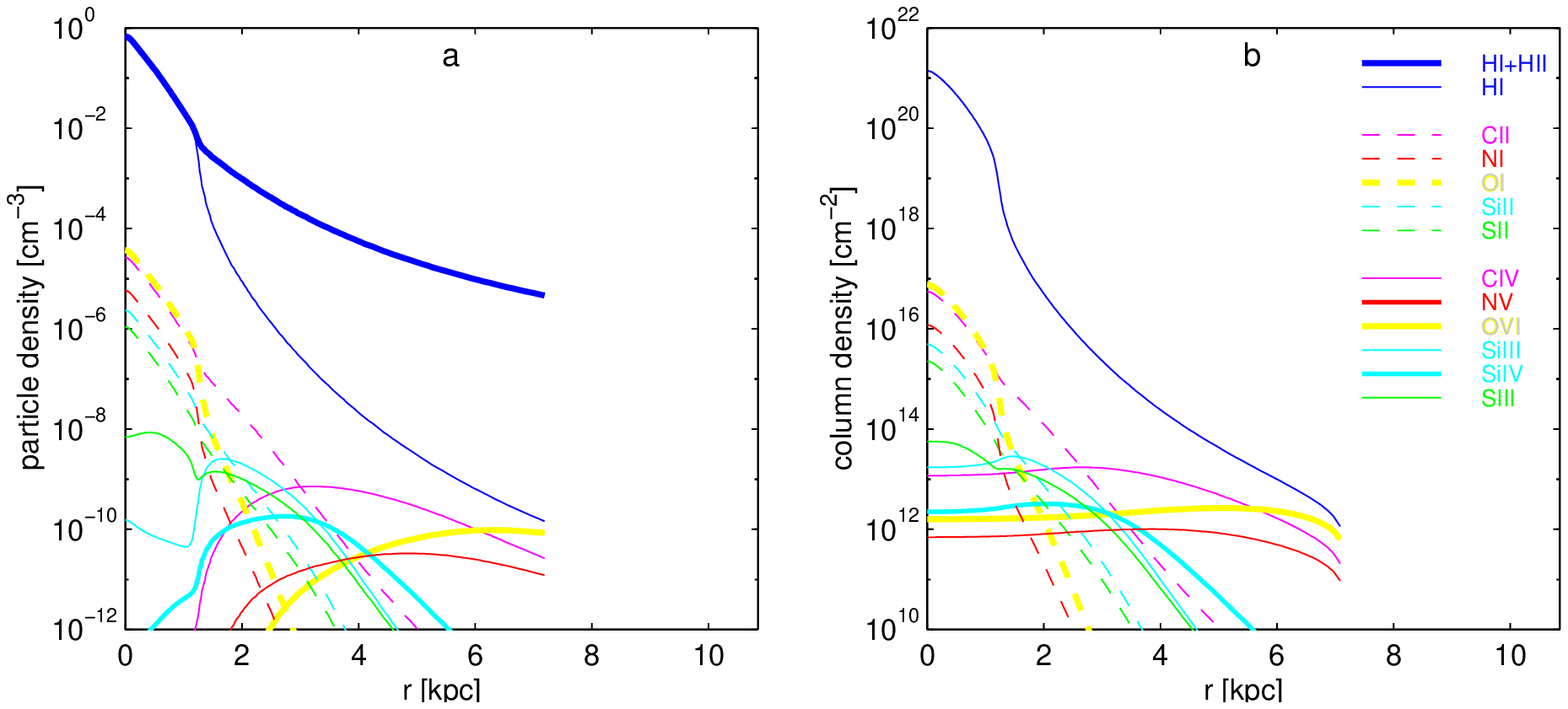}  
\caption{
Model D:
$M_{vir} = 2\times10^9$~M$_{\odot}$ median halo, 
$M_{gas} = 1.6\times10^7$~M$_{\odot}$,
$0.1$~solar metallicity,
standard radiation field, 
$P_{\rm HIM}/k_{B} = 0.1$~cm$^{-3}$~K.}
\label{bigdwarf}
\end{figure}

\begin{figure}
\plotone{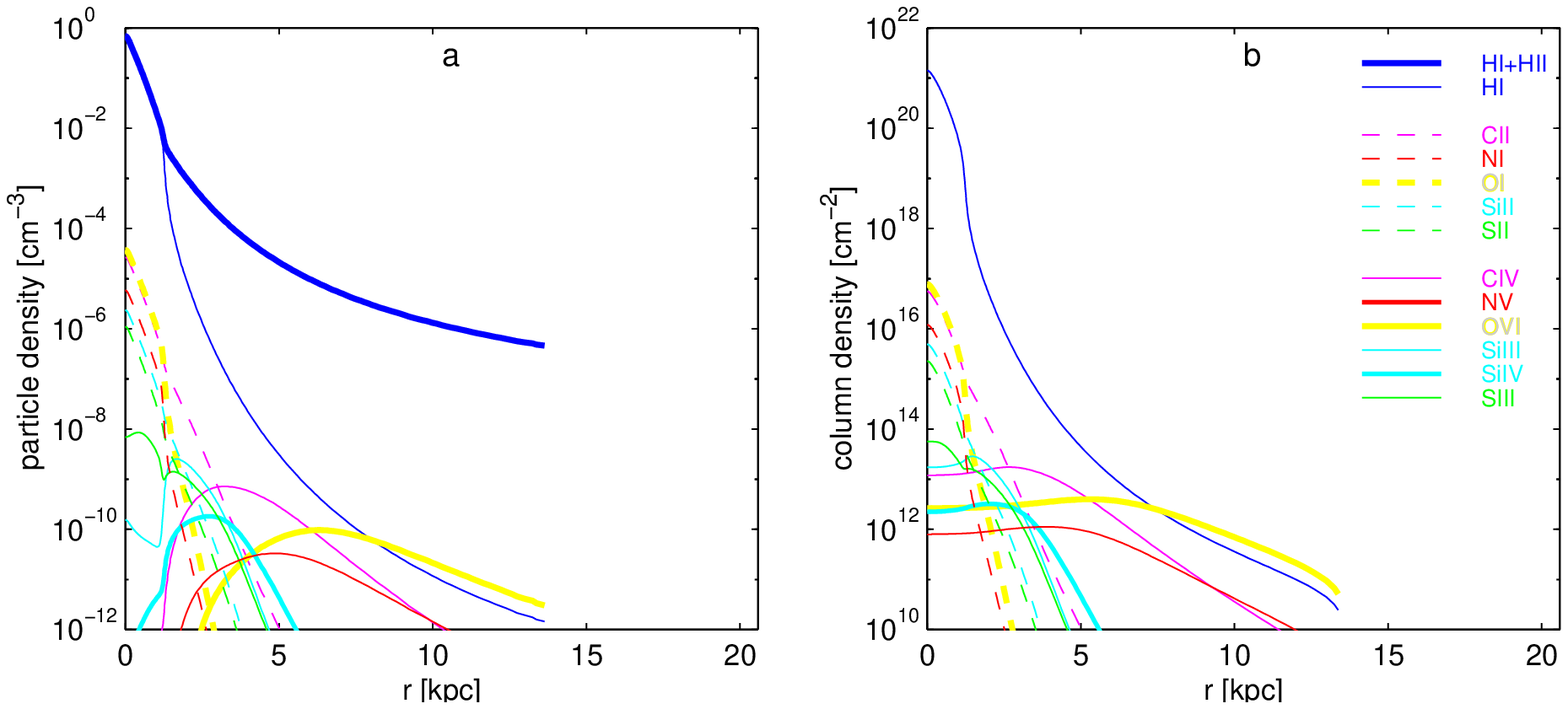}  
\caption{
Model E:
$M_{vir} = 2\times10^9$~M$_{\odot}$ median halo, 
$M_{gas} = 1.7\times10^7$~M$_{\odot}$,
$0.1$~solar metallicity,
standard radiation field, 
$P_{\rm HIM}/k_{B} = 0.01$~cm$^{-3}$~K.}
\label{hugedwarf}
\end{figure}

\begin{figure}
\plotone{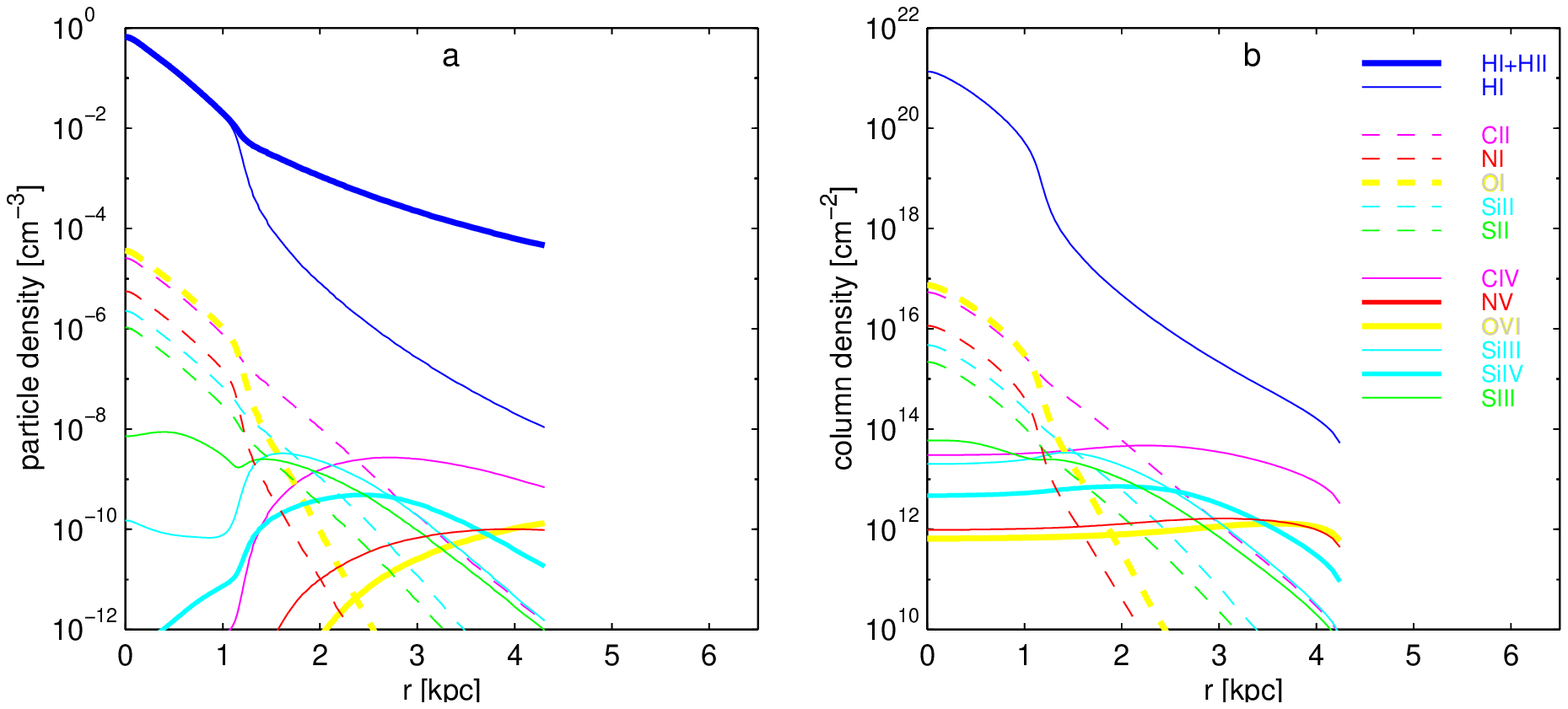}  
\caption{
Model F:
$M_{vir} = 2\times10^9$~M$_{\odot}$ median halo, 
$M_{gas} = 1.6\times10^7$~M$_{\odot}$,
$0.1$~solar metallicity,
maximal radiation field, 
$P_{\rm HIM}/k_{B} = 1$~cm$^{-3}$~K.}
\label{maxspec}
\end{figure}

\begin{figure}
\includegraphics[width=8cm,height=18cm]{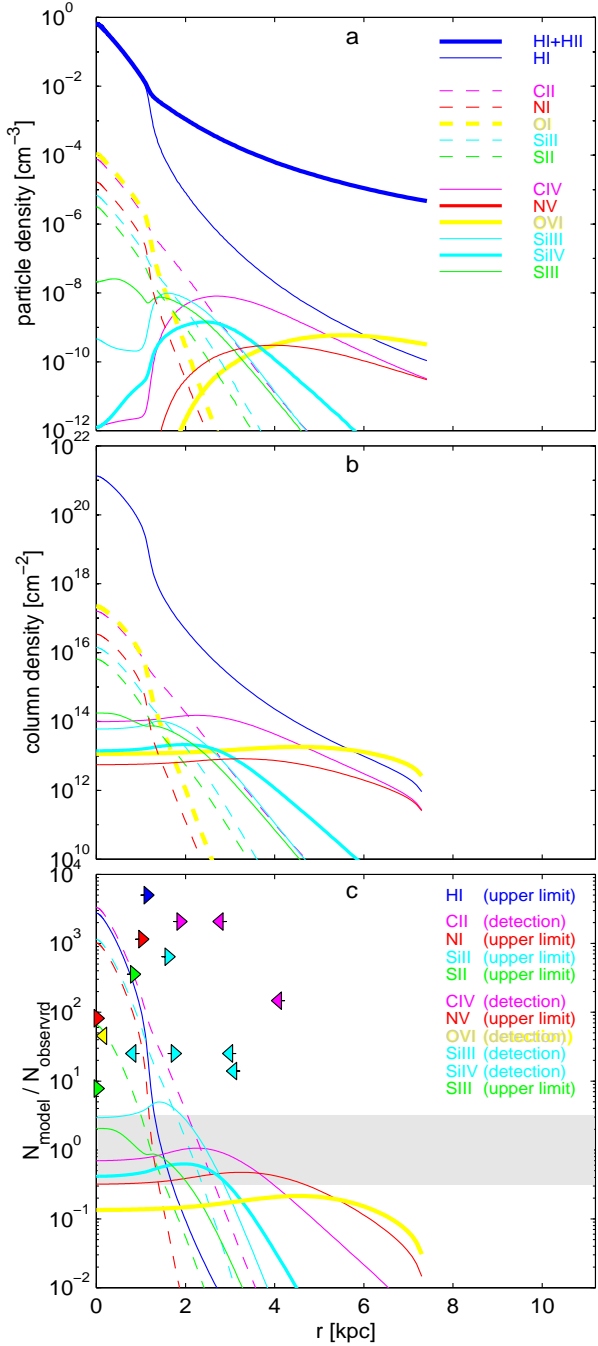}  
\caption{
Model G:
$M_{vir} = 2\times10^9$~M$_{\odot}$ median halo, 
$M_{gas} = 1.6\times10^7$~M$_{\odot}$,
$0.3$~solar metallicity,
maximal radiation field, 
$P_{\rm HIM}/k_{B} = 0.1$~cm$^{-3}$~K.
As before, panel (a) shows the volume densities as a function of the
distance from the cloud center, 
and panel (b) shows the projected column densities.
Panel (c) displays, for each ion,  the ratio of the calculated 
model-G columns to the 
observed columns of the $-300$~km~s$^{-1}$ component towards Mrk~509
(Collins et al.~2004; see Table \ref{collins}).
The legend indicates species with measured columns, 
upper limits or lower limits.
In the shaded area the 
calculated and observed values 
agree to within $0.5$ dex.
For each of the species, the arrows 
indicate the range of radii (between the arrowheads) where the 
model columns are within 0.5 dex of the measured values or limits.}
\label{big0.3max}
\end{figure}

\begin{figure}
\plotone{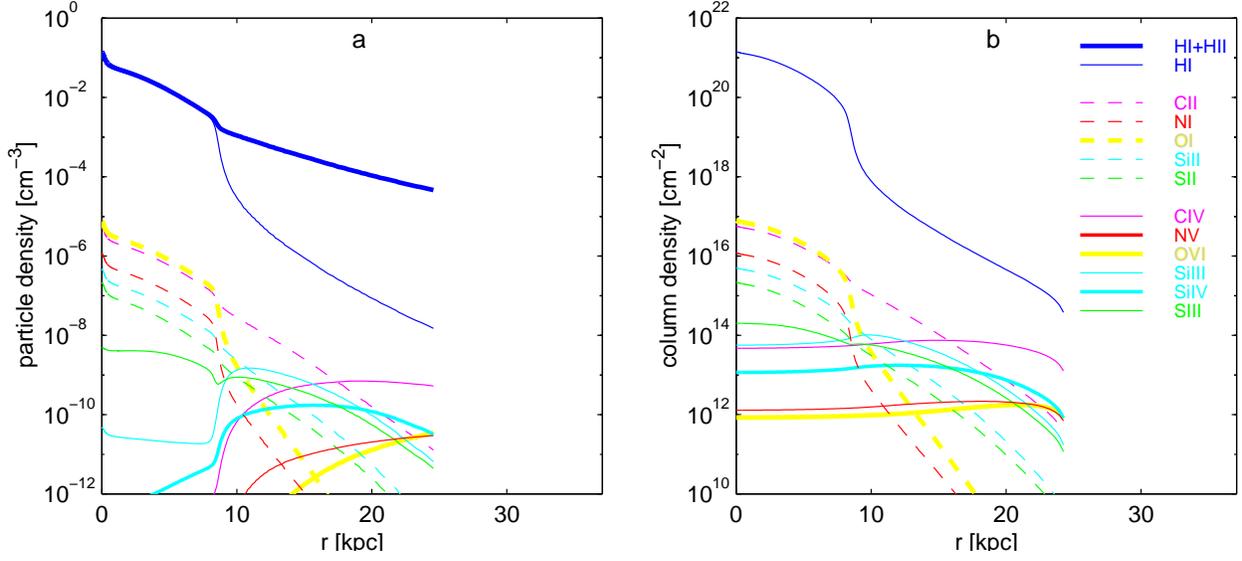}  
\caption{
Model H:
$M_{vir} = 6.2\times10^9$~M$_{\odot}$ $-5\sigma$ halo, 
$M_{gas} = 1.3\times10^9$~M$_{\odot}$,
$0.1$~solar metallicity,
standard radiation field, 
$P_{\rm HIM}/k_{B} = 1$~cm$^{-3}$~K.}
\label{-5sig}
\end{figure}

\begin{figure}
\plotone{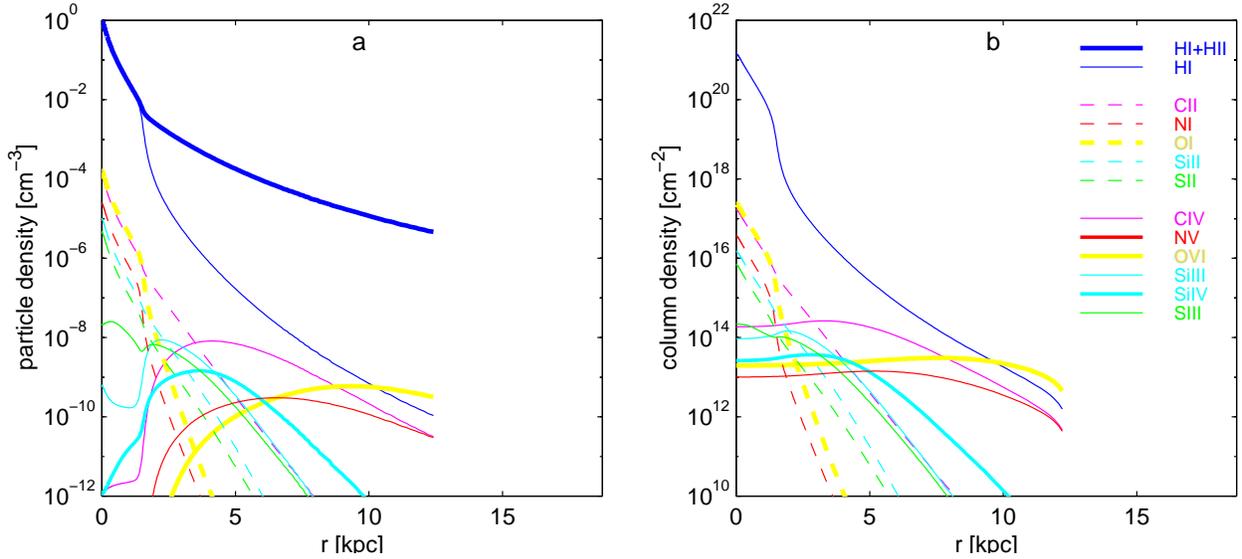}  
\caption{
Model I:
$M_{vir} = 4.9\times10^9$~M$_{\odot}$ $-4\sigma$ NFW halo, 
$M_{gas} = 3.1\times10^7$~M$_{\odot}$,
$0.3$~solar metallicity,
maximal radiation field, 
$P_{\rm HIM}/k_{B} = 0.1$~cm$^{-3}$~K.}
\label{-4sigNFW}
\end{figure}

\begin{figure}
\plotone{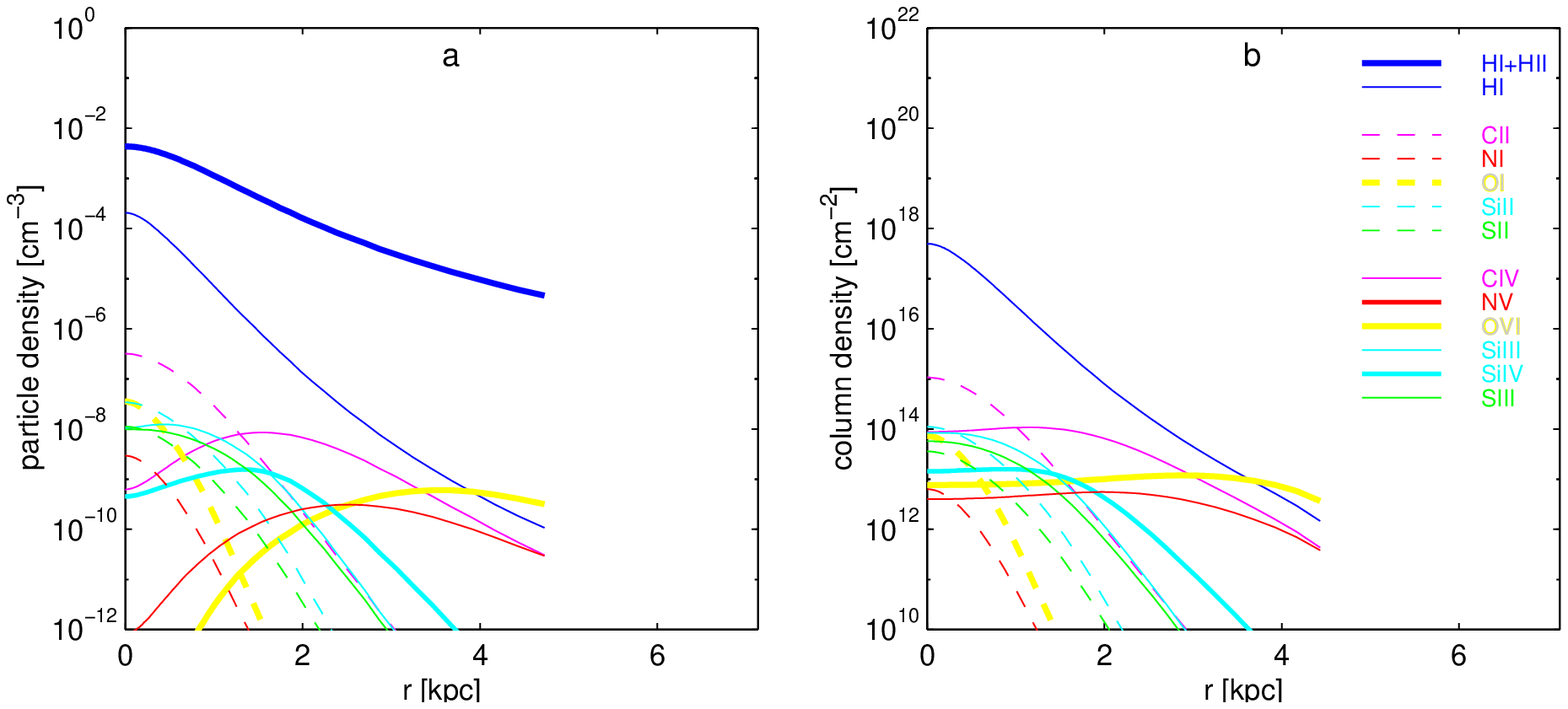}  
\caption{
Model J:
$M_{vir} = 2\times10^9$~M$_{\odot}$ median halo, 
$M_{gas} = 9.6\times10^5$~M$_{\odot}$,
$0.3$~solar metallicity,
maximal radiation field, 
$P/k_{B} = 0.1$~cm$^{-3}$~K.}
\label{bigImax0.3}
\end{figure}

\clearpage



\begin{deluxetable}{lr}
\tablewidth{0pt}
\tablecaption{$0.1$~solar metallicity abundances\label{Holmet}}
\tablehead{
\colhead{Element} & 
\colhead{Abundance}\\
\colhead{} &
\colhead{(X/H)$_{\odot}$} }
\startdata
Carbon   & $3.91\times10^{-5}$  \\
Nitrogen & $8.53\times10^{-6}$  \\
Oxygen   & $5.44\times10^{-5}$  \\
Silicon  & $3.44\times10^{-6}$  \\
Sulfur   & $1.62\times10^{-6}$  \\
\enddata
\tablecomments{
Holweger (2001)
}
\end{deluxetable}


\begin{deluxetable}{lllll}
\tablewidth{0pt}
\tablecaption{Observed column densities - Sembach et al. (1999)\label{sembach}}
\tablehead{
\colhead{ } & 
\colhead{Mrk 509} &
\colhead{ } &
\colhead{PKS 2155-304}  &
\colhead{ }  }
\startdata
           & -283 km s$^{-1}$ &  -228 km s$^{-1}$  &  -256 km s$^{-1}$   &  -140 km s$^{-1}$  \\
\tableline
\ion{H}{1} & $<4.90\times10^{17}$  &  $<4.90\times10^{17}$   &   $<5.37\times10^{17}$   &    $<5.37\times10^{17}$ \\
\ion{C}{2} & $ 4.27^{+2.19}_{-2.22}\times10^{13}$  & $<3.09\times10^{13}$   &              &             \\
\ion{N}{1} & $<3.80\times10^{13}$  &  $<2.45\times10^{13}$   &              &             \\
\ion{Si}{2}& $<1.00\times10^{13}$  &  $<6.92\times10^{12}$   &   $<7.24\times10^{12}$   &    $ (5.37\pm0.14)\times10^{13}$ \\
\ion{S}{2} & $<8.51\times10^{13}$  &  $<6.61\times10^{13}$   &   $<1.58\times10^{14}$   &    $<1.32\times10^{14}$ \\
\\
\ion{C}{4} & $>1.58\times10^{14}$  &  $>4.79\times10^{13}$   &   $(3.16\pm0.47)\times10^{13}$ &    $(3.16\pm0.47)\times10^{13}$  \\
\ion{N}{5} & $<1.51\times10^{13}$  &  $<9.55\times10^{12}$   &   $<1.12\times10^{13}$   &    $<1.38\times10^{13}$ \\
\ion{Si}{3}& $>1.55\times10^{13}$  &  $2.51^{+2.27}_{-2.21}\times10^{12}$    &              &             \\
\ion{Si}{4}& $>2.75\times10^{13}$  &  $<8.51\times10^{12}$   &              &             \\
\ion{S}{3} & $<2.29\times10^{14}$  &  $<2.57\times10^{14}$   &              &             \\
\\
\ion{O}{6} & \multicolumn{2}{c}{$5.75\times10^{13} - 3.02\times10^{14}$} &\multicolumn{2}{c}{$3.71\times10^{12} - 1.10\times10^{14}$} \\
\enddata
\tablecomments{
Sembach et al. (1999) (their table 5) - log column densities for
two velocity components towards Mrk~509, and two 
velocity-components toward PKS~2155-304.
The error estimates and upper limits are the $2\sigma$
estimates (The \ion{H}{1} upper limits are $4\sigma$ estimates).
The lower limits are due to the possibility of unresolved
saturation in the absorption line measurements.
OVI values are from Sembach et al. 2000 and 2003.
}
\end{deluxetable}


\begin{deluxetable}{lllll}
\tablewidth{0pt}
\tablecaption{Observed column densities - Collins et al. (2004)\label{collins}}
\tablehead{
\colhead{ } &
\colhead{Mrk 509} &
\colhead{ } &
\colhead{PKS 2155-304}  &
\colhead{ }  }
\startdata
           & -300 km s$^{-1}$                      &  -240 km s$^{-1}$                      &  -270 km s$^{-1}$                     &  -140 km s$^{-1}$                     \\
\tableline
\ion{H}{1} & $<4.90\times10^{17}$                  &  $<4.90\times10^{17}$                  &  $1.70^{+2.38}_{-0.68}\times10^{15}$  &  $2.34^{+1.55}_{-0.65}\times10^{16}$  \\
\ion{C}{2} & $ 4.79^{+0.97}_{-0.90}\times10^{13}$  &  $<2.09\times10^{13}$                  &  $<1.62\times10^{13}$                 &  $6.92^{+1.79}_{-1.03}\times10^{13}$  \\
\ion{C}{3} & $$                                    &  $$                                    &  $3.31^{+0.24}_{-0.29}\times10^{13}$  &  $7.41^{+0.35}_{-0.33}\times10^{13}$  \\
\ion{N}{1} & $<3.39\times10^{13}$                  &  $<2.40\times10^{13}$                  &  $<2.09\times10^{13}$                 &  $<1.74\times10^{13}$                 \\
\ion{N}{2} & $$                                    &  $$                                    &  $<5.13\times10^{13}$                 &  $1.15^{+0.76}_{-0.46}\times10^{14}$  \\
\ion{Si}{2}& $<1.23\times10^{13}$                  &  $<8.51\times10^{12}$                  &  $<5.89\times10^{12}$                 &  $6.46^{+2.05}_{-1.33}\times10^{12}$  \\
\ion{S}{2} & $<1.00\times10^{14}$                  &  $<7.08\times10^{13}$                  &  $<8.71\times10^{13}$                 &  $<6.46\times10^{13}$                 \\
\\
\ion{C}{4} & $ 1.41^{+0.37}_{-0.21}\times10^{14}$  &  $ 3.39^{+0.41}_{-0.44}\times10^{13}$  &  $3.98^{+0.38}_{-0.27}\times10^{13}$  &  $3.24^{+0.23}_{-0.28}\times10^{13}$  \\
\ion{N}{5} & $<1.74\times10^{13}$                  &  $<1.20\times10^{13}$                  &  $<1.23\times10^{13}$                 &  $<1.20\times10^{13}$                 \\
\ion{O}{6} & $ 8.51^{+0.82}_{-0.93}\times10^{13}$  &  $ 7.76^{+0.75}_{-0.68}\times10^{13}$  &  $3.63^{+0.44}_{-0.47}\times10^{13}$  &  $6.31^{+0.45}_{-0.42}\times10^{13}$  \\
\ion{Si}{3}& $2.04^{+0.15}_{-0.18}\times10^{13}$  &  $ 2.75^{+0.71}_{-0.66}\times10^{12}$  &  $2.14^{+0.55}_{-0.59}\times10^{12}$  &  $(1.45\pm0.07)\times10^{13}$         \\
\ion{Si}{4}& $ 3.39^{+1.86}_{-0.88}\times10^{13}$  &  $<3.09\times10^{12}$                  &  $1.82^{+0.52}_{-0.56}\times10^{12}$ &  $5.50^{+0.96}_{-0.92}\times10^{12}$  \\
\ion{S}{3} & $<8.51\times10^{13}$                  &  $<6.03\times10^{13}$                  &  $<4.90\times10^{13}$                 &  $<4.79\times10^{13}$                 \\
\enddata
\tablecomments{
Column densities for the velocity components towards 
Mrk~509, and PKS~2155-304.
The error estimates are the $1\sigma$ estimates.
The upper limits are $3\sigma$ limits 
(The \ion{H}{1} upper limits are $4\sigma$ estimates).
The quoted \ion{C}{2}, \ion{C}{4}, \ion{Si}{2}, and \ion{Si}{4} columns are
based on observations of the $\lambda 1036.337$, $\lambda1548.203$, 
$\lambda 1193.290$, and $\lambda 1393.755$ lines of these ions respectively.}
\end{deluxetable}


\begin{deluxetable}{lllllllll}
\tablewidth{0pt}
\rotate
\tablecaption{Input Parameters \label{models}}
\tablehead{
\colhead{Model} & 
\colhead{$M_{\rm vir}$} &
\colhead{$x_{\rm vir}$} &
\colhead{$v_s$} &
\colhead{$r_s$} &
\colhead{$M_{\rm WM}$}  &
\colhead{$P_{\rm HIM}$} &
\colhead{radiation field} &
\colhead{metallicity} \\
\colhead{ } & 
\colhead{$[M_{\odot}]$} &
\colhead{ } &
\colhead{[km~s$^{-1}$]} &
\colhead{[kpc]} &
\colhead{$[M_{\odot}]$} &
\colhead{[cm$^{-3}$~K]} &
\colhead{ } &
\colhead{[solar]} }
\startdata
\multicolumn{2}{l}{\underline{Pressure-confined:}} &&&        &                 &        &          &       \\
A & $1.0\times10^8$ & $32.5$                  & $12$ & $0.4$  & $1.1\times10^6$ & $50.0$ & Standard & $0.1$ \\
B & $3.2\times10^8$ & $8.15$ ($-4\sigma$)     & $12$ & $2.2$  & $7.1\times10^7$ & $20.0$ & Standard & $0.1$ \\
\\
\multicolumn{2}{l}{\underline{Gravitationally-confined:}} &&& &                 &        &          &       \\
C & $2.0\times10^9$ & $25.5$                  & $30$ & $1.3$  & $1.6\times10^7$ & $1.0$  & Standard & $0.1$ \\
D & $2.0\times10^9$ & $25.5$                  & $30$ & $1.3$  & $1.6\times10^7$ & $0.1$  & Standard & $0.1$ \\
E & $2.0\times10^9$ & $25.5$                  & $30$ & $1.3$  & $1.7\times10^7$ & $0.01$ & Standard & $0.1$ \\
F & $2.0\times10^9$ & $25.5$                  & $30$ & $1.3$  & $1.6\times10^7$ & $1.0$  & Maximal  & $0.1$ \\
G & $2.0\times10^9$ & $25.5$                  & $30$ & $1.3$  & $1.6\times10^7$ & $0.1$  & Maximal  & $0.3$ \\
H & $6.2\times10^9$ & $4.7$  ($-5\sigma$)     & $30$ & $10.2$ & $1.3\times10^9$ & $1.0$  & Standard & $0.1$ \\ 
I & $4.9\times10^9$ & $6.6$  ($-4\sigma$ NFW) & $30$ & $6.7$  & $3.1\times10^7$ & $0.1$  & Maximal  & $0.3$ \\
J & $2.0\times10^9$ & $25.5$                  & $30$ & $1.3$  & $9.6\times10^5$ & $0.1$  & Maximal  & $0.3$ \\
\enddata
\end{deluxetable}


\begin{deluxetable}{llll}
\tablewidth{0pt}
\tablecaption{Results for pressure-confined models \label{ceres}}
\tablehead{
\colhead{}  & 
\colhead{}  & 
\colhead{A} & 
\colhead{B}
}
\startdata
$N_{HI}$ (peak)             & [cm$^{-2}$]   & $4.9\times10^{19}$ & $9.7\times10^{19}$ \\
$r_{\rm WNM}$               & [kpc]         & $0.67$             & $3.40$              \\
$M_{\rm WNM}$               & [M$_\odot$]   & $3.8\times10^5$    & $2.0\times10^7$     \\
\\
$r_{\rm W\slash H}$         & [kpc]         & $1.30$             & $7.15$              \\
$U_{\rm W\slash H}$         &               & $1.3\times10^{-4}$ & $3.5\times10^{-4}$  \\
$M_{\rm WIM}$               & [M$_\odot$]   & $7.2\times10^5$    & $5.1\times10^7$     \\
\\
\tableline
\\
$N_{CIV}$ (peak)            & [cm$^{-2}$]   & $2.6\times10^{11}$ & $4.6\times10^{12}$ \\
Projected radius at peak $N_{CIV}$ &[kpc]   & $1.1$              & $5.7$               \\
\\
\multicolumn{2}{l}{\underline{Column densities [cm$^{-2}$] at peak $N_{CIV}$:}} & &     \\
\\

\ion{H}{1}                  &               & $3.7\times10^{17}$ & $3.6\times10^{17}$ \\
\ion{C}{2}                  &               & $3.4\times10^{14}$ & $5.8\times10^{14}$ \\
\ion{N}{1}                  &               & $2.7\times10^{12}$ & $1.9\times10^{12}$ \\
\ion{O}{1}                  &               & $2.2\times10^{13}$ & $1.6\times10^{13}$ \\
\ion{Si}{2}                 &               & $2.6\times10^{13}$ & $4.1\times10^{13}$ \\
\ion{S}{2}                  &               & $1.1\times10^{13}$ & $1.7\times10^{13}$ \\
\\
\ion{N}{5}                  &               & $9.5\times10^{8}$  & $4.2\times10^{10}$ \\
\ion{O}{6}                  &               & $3.7\times10^{6}$  & $8.7\times10^{8}$  \\
\ion{Si}{3}                 &               & $1.3\times10^{13}$ & $4.9\times10^{13}$ \\
\ion{Si}{4}                 &               & $3.7\times10^{11}$ & $3.3\times10^{12}$ \\
\ion{S}{3}                  &               & $7.2\times10^{12}$ & $2.6\times10^{13}$ \\
\enddata
\tablecomments{\small{Basic model parameters
and column densities of the various metal species at the location
of \ion{C}{4} peak column, for pressure-confined CHVC mini-halos.}
}
\end{deluxetable}

\clearpage
\thispagestyle{empty}

\begin{deluxetable}{llllllllllll}
\label{}
\tablewidth{0pt}
\rotate
\tabletypesize{\footnotesize}
\tablecaption{Results for gravitationally-confined models \label{dwarfres}}
\tablehead{
\colhead{}              & 
\colhead{}              & 
\colhead{C} &
\colhead{D} & 
\colhead{E} & 
\colhead{F} & 
\colhead{G} &
\colhead{H} &
\colhead{I} &
\colhead{J} &
}
\startdata
$N_{HI}$ (peak)          &  [cm$^{-2}$]      &  $1.4\times10^{21}$  &  $1.4\times10^{21}$  &  $1.4\times10^{21}$  &  $1.4\times10^{21}$  &  $1.3\times10^{21}$  & $1.4\times10^{21}$ & $1.4\times10^{21}$ & $5\times10^{17}$   \\
$r_{\rm WNM}$            &  [kpc]            &  $1.2$               &  $1.2$               &  $1.2$               &  $1.2$               &  $1.2$               & $8.5$              & $1.5$              & $0$                \\
$M_{\rm WNM}$            &  [M$_\odot$]      &  $1.2\times10^7$     &  $1.2\times10^7$     &  $1.2\times10^7$     &  $1.2\times10^7$     &  $1.1\times10^7$     & $8.5\times10^8$    & $1.4\times10^7$    & $0$                \\
\\
$r_{\rm W\slash H}$      &  [kpc]            &  $4.2$               &  $7.7$               &  $13.6$              &  $4.3$               &  $7.42$              & $24.6$             & $12.4$             & $4.7$              \\
$U_{\rm W\slash H}$      &                   &  $8.9\times10^{-3}$  &  $9.0\times10^{-2}$  &  $0.90$              &  $1.6\times10^{-2}$  &  $0.15$              & $8.8\times10^{-3}$ & $0.15$             & $0.16$             \\
$M_{\rm WIM}$            &  [M$_\odot$]      &  $4\times10^6$       &  $4\times10^6$       &  $4\times10^6$       &  $4\times10^6$       &  $5\times10^6$       & $4\times10^8$      & $1.7\times10^7$    & $9.6\times10^5$    \\
\\
\tableline
\\
$N_{CIV} (peak)$        & [cm$^{-2}$]        & $1.3\times10^{13}$   & $1.7\times10^{13}$   & $1.7\times10^{13}$   & $4.7\times10^{13}$   & $1.5\times10^{14}$   & $7.5\times10^{13}$ & $2.6\times10^{14}$ & $1.1\times10^{14}$ \\
Projected radius at peak $N_{CIV}$ &[kpc]    & $2.6$                & $2.6$                & $2.6$                & $2.3$                & $2.3$                & $15.0$             & $3.3$              & $1.1$              \\
\\
\multicolumn{3}{l}{\underline{Column densities [cm$^{-2}$] at peak $N_{CIV}$:}}  & & & & & & & \\
\\
HI                       &                   &  $7.3\times10^{15}$  &  $5.9\times10^{15}$  &  $5.9\times10^{15}$  &  $1.8\times10^{16}$  &  $1.8\times10^{16}$  & $4.0\times10^{16}$ & $3.0\times11^{16}$ & $1.7\times10^{16}$ \\
CII                      &                   &  $2.1\times10^{13}$  &  $1.7\times10^{13}$  &  $1.7\times10^{13}$  &  $2.1\times10^{13}$  &  $6.5\times10^{13}$  & $1.1\times10^{14}$ & $1.1\times10^{14}$ & $6.6\times10^{13}$ \\
NI                       &                   &  $6.9\times10^{9}$   &  $4.4\times10^{9}$   &  $4.5\times10^{9}$   &  $5.9\times10^{9}$   &  $1.8\times10^{10}$  & $3.4\times10^{10}$ & $3.0\times10^{10}$ & $2.3\times10^{10}$ \\
OI                       &                   &  $3.7\times10^{10}$  &  $2.3\times10^{10}$  &  $2.3\times10^{10}$  &  $3.8\times10^{10}$  &  $1.2\times10^{11}$  & $1.9\times10^{11}$ & $2.0\times10^{11}$ & $1.6\times10^{11}$ \\
SiII                     &                   &  $1.1\times10^{12}$  &  $8.3\times10^{11}$  &  $8.4\times10^{11}$  &  $2.1\times10^{12}$  &  $6.2\times10^{12}$  & $5.6\times10^{12}$ & $1.0\times10^{13}$ & $6.4\times10^{12}$ \\
SII                      &                   &  $4.8\times10^{11}$  &  $3.7\times10^{11}$  &  $3.7\times10^{11}$  &  $6.1\times10^{11}$  &  $1.8\times10^{12}$  & $2.4\times10^{12}$ & $3.1\times10^{12}$ & $1.9\times10^{12}$ \\
\\
NV                       &                   &  $3.4\times10^{11}$  &  $9.0\times10^{11}$  &  $1.0\times10^{12}$  &  $1.4\times10^{12}$  &  $7.1\times10^{12}$  & $2.0\times10^{12}$ & $1.2\times10^{13}$ & $4.7\times10^{12}$ \\
OVI                      &                   &  $2.1\times10^{11}$  &  $1.9\times10^{12}$  &  $3.0\times10^{12}$  &  $8.6\times10^{11}$  &  $1.3\times10^{13}$  & $1.3\times10^{12}$ & $2.2\times10^{13}$ & $8.2\times10^{12}$ \\
SiIII                    &                   &  $7.0\times10^{12}$  &  $6.1\times10^{12}$  &  $6.1\times10^{12}$  &  $1.1\times10^{13}$  &  $3.3\times10^{13}$  & $3.7\times10^{13}$ & $5.6\times10^{13}$ & $3.0\times10^{13}$ \\
SiIV                     &                   &  $2.8\times10^{12}$  &  $2.9\times10^{12}$  &  $2.9\times10^{12}$  &  $6.8\times10^{12}$  &  $2.0\times10^{13}$  & $1.6\times10^{13}$ & $3.5\times10^{13}$ & $1.5\times10^{13}$ \\
SIII                     &                   &  $3.7\times10^{12}$  &  $3.2\times10^{12}$  &  $3.3\times10^{12}$  &  $5.7\times10^{12}$  &  $1.7\times10^{13}$  & $2.0\times10^{13}$ & $2.9\times10^{13}$ & $1.6\times10^{13}$ \\
\enddata
\tablecomments{
Same as Table \ref{ceres} but for dwarf-galaxy scale mini-halos.
}
\end{deluxetable}


\end{document}